\documentclass[a4paper,11pt]{article}
\usepackage[a4paper,left=2.73cm,right=2.7cm,top=3cm,bottom=3.5cm]{geometry}

\usepackage{amsfonts,amsmath,amssymb,amsthm,tabstackengine}
\usepackage[usenames, dvipsnames]{color}
\usepackage{array}

\usepackage{color, colortbl}
\definecolor{Gray}{gray}{0.95}

\usepackage[colorlinks=true,linktocpage=true,linkcolor=red,citecolor=blue]{hyperref}
\usepackage{graphicx}

\usepackage{comment}

\numberwithin{equation}{section}

\begin{document}

\begin{titlepage}

\thispagestyle{empty}

\begin{center}

\mbox{{\LARGE \textbf{$\mathcal{N}=2$ $\,\textrm{CFT}_{3}\textrm{'s}\,$ from $\,\mathcal{N} = 4\,$ gauged supergravity}}}

\vspace{40pt}

{\large \bf Miguel Chamorro-Burgos}$\,^{a}$\,  \large{,}  \,   {\large \bf Adolfo Guarino}$\,^{a, b}$\, \large{and}  \, {\large \bf Colin Sterckx}$\,^{c, a}$ 
		
\vspace{25pt}
		
$^a$\,{\normalsize  
Departamento de F\'isica, Universidad de Oviedo,\\
Avda. Federico Garc\'ia Lorca 18, 33007 Oviedo, Spain.}
\\[7mm]

$^b$\,{\normalsize  
Instituto Universitario de Ciencias y Tecnolog\'ias Espaciales de Asturias (ICTEA) \\
Calle de la Independencia 13, 33004 Oviedo, Spain.}
\\[7mm]

$^c$\,{\normalsize  
Universit\'e Libre de Bruxelles (ULB) and International Solvay Institutes,\\
Service  de Physique Th\'eorique et Math\'ematique, \\
Campus de la Plaine, CP 231, B-1050, Brussels, Belgium.}
\\[10mm]


\abstract{
\noindent We use holography and four-dimensional $\,\mathcal{N}=4\,$ gauged supergravity to collect evidence for a large class of interconnected three-dimensional $\,\mathcal{N}=2\,$ conformal field theories. On the gravity side, we construct a one-parameter family of $\,{\textrm{ISO}(3) \times \textrm{ISO}(3)}$ gaugings of half-maximal supergravity containing a rich structure of $\,\mathcal{N}=2\,$ AdS$_{4}$ solutions at fixed radius. By looking at excitations around these AdS$_{4}$ solutions, the spectrum of low lying operators in the dual $\,\mathcal{N}=2\,$ CFT$_{3}$'s is computed and further arranged into $\,\mathfrak{osp}(2|4)$ supermultiplets. Upon suitable removal of gauge redundancies, we identify the Zamolodchikov metric on the conformal manifold dual to the AdS$_{4}$ moduli space, and recover previous results in the S-fold literature. Two special points of $\,\mathcal{N}=4\,$ supersymmetry enhancement occur. While one describes an S-fold CFT$_{3}$ dual to a non-geometric type IIB twisted compactification, the string-theoretic realisation of the other, if any, is still lacking.
}

\end{center}

\end{titlepage}

\tableofcontents

\hrulefill
\vspace{10pt}

\section{Introduction}

Conformal field theories in three space-time dimensions (CFT$_{3}$'s) have played and continue to play a central role in string and M-theory. When endowed with enough supersymmetry, various non-renormalisation theorems have been stated making superconformal field theories (SCFT$_{3}$'s) very interesting from the viewpoint of the AdS$_{4}$/CFT$_{3}$ correspondence. Interacting superconformal field theories in three space-time dimensions are special because gauge fields are described by (an up to $\,\mathcal{N}=3\,$ supersymmetric extension of) the topological Chern--Simons (CS) Lagrangian\footnote{The standard $\,F^2\,$ term is dimension $\,4\,$ and therefore is not compatible with scale invariance.}
\begin{equation}
\mathcal{L}_\textrm{CS} = \frac{k}{4 \pi} \textrm{Tr} \left[ \epsilon^{\mu\nu\rho} \left( A_{\mu} \partial_{\nu} A_{\rho} + \tfrac{2}{3} A_{\mu} A_{\nu} A_{\rho}  \right)  \right] \ ,
\end{equation}
where $\,k \in \mathbb{Z}\,$ is the so-called CS level. It was then shown in \cite{Schwarz:2004yj} that gauge fields can be consistently coupled to matter fields (scalars and fermions) carrying new degrees of freedom in a way compatible with $\,\mathcal{N}=1,2\,$ supersymmetry as well as with conformal symmetry. The result is a superconformal CS-matter theory with $\,\textrm{OSp}(\mathcal{N}|4)\,$ superconformal symmetry.

\subsection*{CS-matter theories in three dimensions and holography}

One of the motivations to investigate superconformal CS-matter theories in three dimensions was to provide a holographic dual to the maximally supersymmetric $\,\textrm{AdS}_{4} \times \textrm{S}^7\,$ solution of eleven-dimensional (11D) supergravity that describes a stack of $\,N\,$ M2-branes on flat space. An important step towards constructing such an $\,\mathcal{N}=8\,$ CFT$_{3}$ was the BLG theory \cite{Bagger:2006sk,Bagger:2007jr,Bagger:2007vi,Gustavsson:2007vu} with gauge group $\,\textrm{G}=\textrm{SO}(4) \sim \textrm{SU}(2)_{k} \times \textrm{SU}(2)_{-k}\,$ and where $\,k \in \mathbb{Z}\,$ is the CS level of the corresponding gauge factor (see \cite{Bandres:2008vf} for a study of the realisation of the $\,\textrm{OSp}(8|4)\,$ superconformal symmetry in the BLG theory). However, there was no large $\,N\,$ limit to be taken on the BLG theory and, therefore, no gravitational interpretation of the CS level $\,k\,$. These questions cleared up with the advent of the $\,\mathcal{N}=6\,$ ABJM theory \cite{Aharony:2008ug}: a CS-matter theory with gauge group $\,\textrm{G}=\textrm{U}(N)_{k} \times \textrm{U}(N)_{-k}\,$. In the large $\,N\,$ limit, it describes an $\,\textrm{AdS}_{4} \times \textrm{S}^7/\mathbb{Z}_{k}\,$ orbifold solution of 11D supergravity and is conjectured to feature $\,{\mathcal{N}=8}\,$ supersymmetry enhancement at low CS levels $\,k=1,2\,$ (see \cite{Benna:2008zy} for a generalisation of ABJM theory and its geometrical realisation in 11D supergravity). Other interesting SCFT$_{3}$'s with $\,\mathcal{N}=1,2,3\,$ have appeared in the context of ten-dimensional (10D) massive type IIA supergravity with gauge groups $\,\textrm{G}=\textrm{SU}(N)_{k_{1}} \times \textrm{SU}(N)_{k_{2}}\,$ \cite{Gaiotto:2009mv} and $\,\textrm{G}=\textrm{SU}(N)_{k}\,$ \cite{Guarino:2015jca} including charged matter fields transforming in different representations of $\,\textrm{G}\,$. On the gravity side, they correspond to $\,\textrm{AdS}_{4} \times \mathbb{CP}_{3}\,$ and $\,\textrm{AdS}_{4} \times \textrm{S}^{6}\,$ solutions of massive IIA supergravity and the Romans mass parameter is related to the CS levels $\,k_{1}+k_{2} \neq 0\,$ and $\,k\,$, respectively.

SCFT$_{3}$'s have also been extensively investigated in the context of type IIB string theory. A precursor of these theories was the three-dimensional planar interface in four-dimensional $\,{\mathcal{N} = 4}\,$ super Yang--Mills \cite{Clark:2004sb} dual to the simple non-supersymmetric Janus solution of type IIB supergravity constructed in \cite{Bak:2003jk}. Supersymmetric $\,\mathcal{N}=1,2,4\,$ three-dimensional interfaces were constructed in \cite{DHoker:2006qeo} and shown to admit a superconformal limit. The dual type IIB supersymmetric Janus solutions were constructed in \cite{Clark:2005te,DHoker:2006vfr}, \cite{Bobev:2020fon} and \cite{DHoker:2007zhm}, respectively. Motivated by the $\,\mathcal{N}=4\,$ superconformal interface and its type IIB dual of the form $\,\textrm{AdS}_{4} \times \textrm{S}^{2} \times \textrm{S}^{2} \times \Sigma\,$ -- with $\Sigma\,$ being a Riemann surface with coordinates $\,\eta \in (-\infty,\infty)\,$ and $\,\alpha \in [0,\frac{\pi}{2}]\,$ having the topology of the infinite strip --, a general class of half-maximal $\,\mathcal{N}=4\,$ SCFT$_{3}$'s was put forward in \cite{Gaiotto:2008sd,Gaiotto:2008ak} for which brane setups were provided in terms of D$3$-branes suspended between NS$5$-branes and D$5$-branes \cite{Hanany:1996ie}. These brane setups are dual to linear quivers which can be made circular by taking the $\,\eta\,$ direction to be an $\,\textrm{S}^{1}$, namely, by changing the topology of $\Sigma\,$ to that of the annulus. In this context, mirror symmetry \cite{Intriligator:1996ex} between two theories is interpreted as an equivalence relation between their brane setups: D5-branes and NS5-branes are exchanged by the inversion element $\,\mathcal{S} \in \textrm{SL}(2,\mathbb{Z})$. This can be generalised to other elements of $\,\textrm{SL}(2,\mathbb{Z})$. For example, acting $\,k\,$ times with the unit-translation element, \textit{i.e.} $\,\mathcal{T}^{k} \in \textrm{SL}(2,\mathbb{Z})\,$, leaves a D5-brane invariant and transforms an NS$5$-brane into a $\,(1,k)$ $5$-brane. This is interpreted as turning on a CS level $\,k\,$ in the corresponding quiver \cite{Witten:2003ya,Gaiotto:2008ak}. Amongst the class of quivers with a known type IIB brane intersection, the mirror symmetric $\,\textrm{T}[\textrm{U}(N)]\,$ linear quiver plays a prominent role \cite{Gaiotto:2008ak}. This quiver has an infrared (IR) SCFT$_{3}$ fixed point, the $\,\textrm{T}[\textrm{U}(N)]\,$ theory, with a $\,\textrm{U}(N) \times \textrm{U}(N)\,$ global symmetry, although only one $\,\textrm{U}(N)\,$ is present in the ultraviolet (UV) Lagrangian description of the theory. Gauging the diagonal $\,\textrm{U}(N)\,$ subgroup of the $\,\textrm{U}(N) \times \textrm{U}(N)\,$ global symmetry using an $\,\mathcal{N}=4\,$ vector multiplet and turning on a CS coupling $\,k$, it was argued in \cite{Assel:2018vtq} that a novel class of $\,\mathcal{N}=4\,$ CS-matter theories, dubbed S-fold CFT$_{3}$'s, emerges in the IR.

\subsection*{Holographic duals of S-fold CFT$_{3}$'s}

The gravity dual of the simplest $\,\mathcal{N}=4\,$ S-fold CFT$_{3}$ was constructed in \cite{Inverso:2016eet} upon uplift of an AdS$_{4}$ solution originally reported in \cite{Gallerati:2014xra}. It is a type IIB supergravity solution of the form $\,\textrm{AdS}_{4} \times \textrm{S}^{1}_{\eta} \times \textrm{S}^{5}\,$ where the $\,\textrm{SO}(6)\,$ isometry group of the $\,\textrm{S}^{5}\,$ is broken to the  $\,\textrm{SO}(3) \times \textrm{SO}(3) \sim \textrm{SO}(4)_{\textrm{R}}\,$ isometry group\footnote{We have attached an R-symmetry label $\,_{\textrm{R}}\,$ to $\,\textrm{SO}(4)_{\textrm{R}}\,$ in order to highlight the $\,\mathcal{N}=4\,$ supersymmetry emerging in the infrared S-fold CFT$_{3}$.} of the $\,\textrm{S}^{2} \times \textrm{S}^{2} \subset \textrm{S}^{5}$. Therefore, the S-fold geometry lies within the class of $\,\textrm{AdS}_{4}  \times \textrm{S}^{2} \times \textrm{S}^{2}  \times \Sigma\,$ (local) solutions of \cite{DHoker:2007zhm} with an annulus topology for $\,\Sigma\,$. However, when looping around the $\,\textrm{S}^{1}_{\eta} \subset \Sigma$, the $\,\mathcal{N}=4\,$ S-fold solution of \cite{Inverso:2016eet} comes along with a non-trivial $\,\textrm{SL}(2,\mathbb{Z})\,$ monodromy of hyperbolic type, \textit{i.e.} $\,J_{k}=-\mathcal{S}\mathcal{T}^{k}$, for the type IIB supergravity fields transforming under S-duality (hence the term S-fold). This $\,\textrm{SL}(2,\mathbb{Z})\,$ monodromy renders the type IIB S-fold solution non-geometric. Interestingly, the $\,\mathcal{N}=4\,$ S-fold of \cite{Inverso:2016eet} was first obtained as a half-maximal AdS$_{4}$ solution ($16$~supercharges) of an effective four-dimensional maximal supergravity ($32$~supercharges) with a dyonic gauging of the group
\begin{equation}
\label{G_max_embedding}
\textrm{G}_{\textrm{max}} = [\textrm{SO}(1,1) \times \textrm{SO}(6)] \ltimes \mathbb{R}^{12} \subset \textrm{E}_{7(7)} \ ,    
\end{equation}
where $\,\textrm{E}_{7(7)}\,$ is the duality group of maximal supergravity \cite{deWit:2005ub}. After the AdS$_{4}$ solution is found in four dimensions, it is straightforward (but tedious) to uplift it to ten dimensions by implementing a generalised Scherk-Schwarz reduction \cite{Hohm:2014qga} of the $\,\textrm{E}_{7(7)}\,$ exceptional field theory ($\textrm{E}_{7(7)}$-ExFT) \cite{Hohm:2013uia}, and then using the dictionary between $\textrm{E}_{7(7)}$-ExFT and type IIB fields \cite{Baguet:2015sma}. This bottom-up procedure has proven very successful and various other type IIB S-folds have been constructed in this manner \cite{Guarino:2019oct,Guarino:2020gfe,Giambrone:2021wsm,Guarino:2022tlw}.

Of particular interest for this work is the $\,\mathcal{N}=2\,$ S-fold with $\,\textrm{SU}(2) \times \textrm{U}(1)_{\textrm{R}}\,$ symmetry put forward in \cite{Guarino:2020gfe}. Both the $\,\mathcal{N}=4\,$ and $\,\mathcal{N}=2\,$ S-folds have the same AdS$_{4}$ radius and were shown in \cite{Bobev:2021yya} to be connected by a scalar modulus $\,\varphi \in \mathbb{R}\,$, namely, a non-compact flat direction in the scalar potential of the maximal gauged supergravity. In light of the AdS$_{4}$/CFT$_{3}$ correspondence, these two solutions are dual to two different points in a conformal manifold of $\,\mathcal{N}=2\,$ S-fold CFT$_{3}$'s with the scalar modulus $\,\varphi\,$ being dual to an exactly marginal deformation \cite{Giambrone:2021zvp,Bobev:2021yya}. Such a conformal manifold of $\,\mathcal{N}=2\,$ S-fold CFT$_{3}$'s was shown in \cite{Bobev:2021yya} to actually be two-dimensional, with the additional marginal deformation being dual to another flat direction $\,\chi\,$ of the scalar potential originally identified in \cite{Guarino:2020gfe}. The Zamolodchikov metric on the conformal manifold (CM) of $\,\mathcal{N}=2\,$ CFT$_{3}$'s was computed in \cite{Bobev:2021yya} using holography and reads
\begin{equation}
\label{CM_metric_intro}
ds_{\textrm{CM}}^2 = \dfrac{1+2 \varphi^2}{2(1+\varphi^2)^2} \big( d\varphi^2 + (1+\varphi^2) \, d\chi^2 \big) \ .
\end{equation}
The type IIB uplift of the modulus $\,\chi\,$ in the AdS$_{4}$ solutions at $\,\varphi=0\,$ ($\mathcal{N}=2\,$ S-fold) and $\,\varphi=1\,$ ($\mathcal{N}=4\,$ S-fold) showed that $\,\chi\,$ is a compact parameter \cite{Giambrone:2021zvp,Guarino:2021kyp}. However, the apparent non-compactness of the modulus $\,\varphi\,$ \cite{Bobev:2021yya,Cesaro:2021tna} remains a puzzle and poses some challenges to the CFT distance conjecture \cite{Perlmutter:2020buo}. Further insights on this issue could come from the ten-dimensional uplift of the $\,\varphi\,$ modulus which has not been worked out yet. Lastly, $\,\mathcal{N}=1\,$ \cite{Bobev:2021rtg} as well as non-supersymmetric \cite{Giambrone:2021wsm} conformal manifolds of S-fold CFT$_{3}$'s have also been investigated.

\subsection*{Plan of this work}

We will continue using the effective four-dimensional supergravity approach to holographically explore the landscape of $\,\mathcal{N}=2\,$ CFT$_{3}$'s. This approach allows to compute the conformal dimensions of all the operators in the $\,\mathcal{N}=2\,$ CFT$_{3}$'s dual to supergravity excitations of the corresponding AdS$_{4}$ solutions. According to the standard AdS$_{4}$/CFT$_{3}$ correspondence, the relation between the normalised mass $\,m L\,$ of a supergravity field of spin $\,[j]\,$ in a given AdS$_{4}$ solution with radius $\,L\,$ and the conformal dimension $\,\Delta\,$ of the dual operator in the CFT$_{3}$ is given by
\begin{equation}
\begin{array}{lll}
[\frac{3}{2}] &:&  m L = \Delta - \frac{3}{2} \ , \\[3mm]
[1] &:&  m^2 L^2 = (\Delta - 2 ) (\Delta - 1 ) \ , \\[3mm]
[\frac{1}{2}] &:&  m L = \Delta - \frac{3}{2}  \ , \\[3mm]
[0] &:&  m^2 L^2 = \Delta (\Delta - 3) \ ,
\end{array}
\end{equation}
with the graviton $\,[j]=[2]\,$ being massless. Moreover, using three-dimensional $\,\mathcal{N}=2\,$ representation theory, the set of operators can be arranged into superconformal multiplets of the  $\,\mathfrak{osp}(2|4)\,$ superconformal algebra.

We will concentrate on the holographic study of $\,\mathcal{N}=2\,$ CFT$_{3}$'s with a potential type~IIB gravity dual of the form $\,\textrm{AdS}_{4} \times \textrm{S}^{2} \times \textrm{S}^{2} \times \Sigma\,$. Firstly, in order to be general but still keep control over the effective supergravity, we will consider four-dimensional supergravity Lagrangians preserving half-maximal supersymmetry ($16$ supercharges). Secondly, in order to be able to recover the conformal manifold of $\,\mathcal{N}=2\,$ S-fold CFT$_{3}$'s constructed in the context of maximal supergravity \cite{Bobev:2021yya}, we will couple the half-maximal supergravity multiplet to \textit{six} vector multiplets so that the field content of the half-maximal supergravity forms a subset of the (unique) field content of maximal supergravity. Thirdly, in order to be compatible with a potential uplift to a type IIB solution of the form $\,\textrm{AdS}_{4} \times \textrm{S}^{2} \times \textrm{S}^{2} \times \Sigma\,$, the gauging must contain an $\,\textrm{SO}(3) \times \textrm{SO}(3)\,$ factor. It becomes then natural to investigate gaugings of the group 
\begin{equation}
\label{G_half-max_embedding}
\textrm{G}_{\textrm{half-max}}= \textrm{ISO}(3) \times \textrm{ISO}(3) \subset \textrm{SL}(2) \times \textrm{SO}(6,6) \ ,  
\end{equation}
with $\,\textrm{ISO}(3) = \textrm{SO}(3) \ltimes \mathbb{R}^{3}\,$, as well as their possible group-theoretical embeddings (\ref{G_half-max_embedding}) into the $\,\textrm{SL}(2) \times \textrm{SO}(6,6)\,$ duality group of half-maximal supergravity coupled to six vector multiplets. The most general such embeddings turns out to depend on \textit{eight} parameters. However, a full analysis of the eight-parameter family of gaugings in (\ref{G_half-max_embedding}) goes beyond the scope of this work.

In this work we will simplify the setup by choosing the same group-theoretical embedding for the two $\,\textrm{ISO}(3)\,$ factors in (\ref{G_half-max_embedding}). This reduces the number of embedding parameters down to four. Turning on three out of the four embedding parameters produces new multi-parametric families of AdS$_{4}$ solutions, all of them with the same AdS$_{4}$ radius. Interestingly, for any value of the three embedding parameters, the AdS$_{4}$ solutions still feature two scalar moduli $\,(\varphi,\chi) \in \mathbb{R}\,$ associated with non-compact flat directions in the scalar potential. By adjusting the embedding parameters and the scalar moduli $\,(\varphi,\chi)\,$ in the AdS$_{4}$ solutions, these can preserve $\,\mathcal{N}=2\,$ ($8$~supercharges), $\,\mathcal{N}=3\,$ ($12$~supercharges) or $\,\mathcal{N}=4\,$ ($16$~supercharges) supersymmetry. Via the gauge/gravity duality, these AdS$_{4}$ solutions are conjectured to be dual to new classes of strongly-coupled $\,\mathcal{N}=2\,$, $\,\mathcal{N}=3\,$ or $\,\mathcal{N}=4\,$ CFT$_{3}$'s provided an embedding in string theory (yet to be worked out) exists.

Our results point at the existence of a web of $\,\mathcal{N}=2\,$ CFT$_{3}$'s containing a special ``line" of $\,\mathcal{N}=3\,$ supersymmetry enhancement which, in turn, contains isolated ``points" where supersymmetry gets enhanced to $\,\mathcal{N}=4\,$. We will characterise this web of $\,\mathcal{N}=2\,$ CFT$_{3}$'s by arranging their set of low lying operators into superconformal multiplet of $\,\mathfrak{osp}(2|4)$, and by further discussing the phenomenon of supermultiplet shortening that occurs at the special loci where supersymmetry gets enhanced to $\,\mathcal{N}=3\,$ or $\,\mathcal{N}=4\,$. Regarding the latter, two isolated points describing $\,\mathcal{N}=4\,$ CFT$_{3}$'s are identified. The first one describes the $\,\mathcal{N}=4\,$ S-fold CFT$_{3}$ of \cite{Assel:2018vtq,Inverso:2016eet} dual to the $\,\mathcal{N}=4\,$ AdS$_{4}$ solution of the maximal supergravity with gauge group (\ref{G_max_embedding}).\footnote{Note that $\,\textrm{G}_{\textrm{half-max}} \subset \textrm{G}_{\textrm{max}}\,$. This makes it possible to rediscover an AdS$_{4}$ solution of a maximal supergravity as an AdS$_{4}$ solution of a half-maximal supergravity provided certain relations between the couplings in the half-maximal supergravity Lagrangian hold (see eq.(\ref{QC_Extra})).} The second one describes a novel $\,\mathcal{N}=4\,$ CFT$_{3}$ -- we will refer to it as the \textit{exotic} $\,\mathcal{N}=4\,$ CFT$_{3}$ -- dual to an $\,\mathcal{N}=4\,$ AdS$_{4}$ solution of a half-maximal supergravity with gauge group (\ref{G_half-max_embedding}).\footnote{This $\,\mathcal{N}=4\,$ AdS$_{4}$ solution is ``maximally" supersymmetric within the theory it belongs to: it preserves the $16$ supercharges of the half-maximal supergravity where it lives.} This AdS$_{4}$ solution appeared originally in \cite{Dibitetto:2011gm} where it was argued to be a non-geometric solution still admitting a locally geometric type IIB description. We now see that it is actually connected to the locally geometric $\,\mathcal{N}=4\,$ S-fold solution of type IIB supergravity \cite{Inverso:2016eet}, at least at the effective four-dimensional supergravity level. As a by-product, we will also present an additional set of generically non-supersymmetric marginal deformations of the exotic $\,\mathcal{N}=4\,$ CFT$_{3}$ which resemble (without being the same) the axion-like deformations of S-folds introduced in \cite{Giambrone:2021zvp,Guarino:2021kyp,Guarino:2021hrc,Guarino:2022tlw}.

The paper is organised as follows. In Section~\ref{sec:N=4_model} we review half-maximal gauged supergravity in four dimensions and present a simple class of $\,\textrm{ISO}(3) \times \textrm{ISO}(3)\,$ gaugings which depends on a specific \textit{deformation} parameter $\,\tilde{\varphi}\,$. In Section~\ref{sec:Z2xZ2_sector} we construct the $\mathbb{Z}_{2}^2$-invariant sector of the theory and present a $\,(\varphi;\tilde{\varphi})$-family of $\,\mathcal{N}=2\,$ AdS$_{4}$ solutions that incorporates the $\,\varphi\,$ modulus dual to one of the two marginal operators spanning the conformal manifold of $\,\mathcal{N}=2\,$ S-fold CFT$_{3}$'s. These AdS$_{4}$ solutions are holographically conjectured to describe a web of $\,\mathcal{N}=2\,$ CFT$_{3}$'s whose spectrum of low lying operators is arranged into superconformal multiplets of $\,\mathfrak{osp}(2|4)$. Supersymmetry as well as flavour symmetry enhancements are discussed together with the corresponding shortening of superconformal multiplets. In Section~\ref{sec:U(1)_sector} we construct the $\textrm{U}(1)_{\textrm{R}}$-invariant sector of the theory in order to also incorporate the modulus $\,\chi\,$ dual to the second marginal operator compatible with $\,\mathcal{N}=2\,$ supersymmetry in the dual CFT$_{3}$'s. We present a $\,(\varphi,\chi;\tilde{\varphi})$-family of $\,\mathcal{N}=2\,$ AdS$_{4}$ solutions that generalises the results of \cite{Bobev:2021yya} to the context of half-maximal supergravity, and arrange the spectrum of low lying operators of the dual $\,\mathcal{N}=2\,$ CFT$_{3}$'s into superconformal multiplets of $\,\mathfrak{osp}(2|4)$. After a suitable treatment of vector fields and gauge redundancies in the $\textrm{U}(1)_{\textrm{R}}$-invariant sector, the Zamolodchikov metric on the conformal manifold of such $\,\mathcal{N}=2\,$ CFT$_{3}$'s is shown to still be (\ref{CM_metric_intro}) for arbitrary values of the half-maximal deformation parameter $\,\tilde{\varphi}\,$. We conclude in Section~\ref{sec:conclusions} with some implications and potential applications of the results presented in this work. Appendix~\ref{app:ISO(3)xISO(3)_general_gauging} discusses more general gaugings of $\,\textrm{ISO}(3) \times \textrm{ISO}(3)\,$ in half-maximal supergravity.

\section{$\textrm{ISO}(3) \times \textrm{ISO}(3)\,$ half-maximal supergravity}
\label{sec:N=4_model}

As stated in the introduction, our starting point is the four-dimensional maximal ($\mathcal{N}=8$) supergravity with gauge group
\begin{equation}
\textrm{G}_{\textrm{max}} = [\textrm{SO}(1,1) \times \textrm{SO}(6)] \ltimes \mathbb{R}^{12} \subset \textrm{E}_{7(7)} \ ,    
\end{equation}
put forward in \cite{Inverso:2016eet}, and shown to accommodate various classes of $\,\textrm{AdS}_{4}\,$ solutions uplifting to S-fold backgrounds of type IIB supergravity \cite{Inverso:2016eet,Guarino:2019oct,Guarino:2020gfe,Giambrone:2021wsm,Guarino:2022tlw}. Following the prescription in \cite{Dibitetto:2011eu}, we will mod out this maximal supergravity by a $\,\mathbb{Z}_{2}\,$ discrete group to produce a very specific half-maximal ($\mathcal{N}=4$) supergravity with gauge group
\begin{equation}
\textrm{G}_{\textrm{half-max}}= \textrm{ISO}(3) \times \textrm{ISO}(3) \subset \textrm{SL}(2) \times \textrm{SO}(6,6) \ .  
\end{equation}
Lastly, we will introduce a specific deformation of such a half-maximal supergravity -- which we parameterise in terms of a continuous parameter $\,\tilde{\varphi} \in \mathbb{R}\,$ -- and characterise it from an algebraic perspective.

\subsection{A crash course on $\,\mathcal{N}=4\,$ gauged supergravity}

We frame our work within the context of half-maximal $\,\mathcal{N}=4\,$ gauged supergravity in four dimensions \cite{Schon:2006kz}. The ungauged theory features a global duality group  $\,\mathcal{G}=\textrm{SL}(2) \times \textrm{SO}(6,n)\,$ where $\,n\,$ is the number of vector multiplets to which the supergravity multiplet is coupled. Promoting a subgroup $\,\textrm{G} \subset \mathcal{G}\,$ from global to local, \textit{i.e.} performing a so-called \textit{gauging}, the duality group $\,\mathcal{G}\,$ is generically broken and only the local gauge symmetry $\,\textrm{G}\,$ is left in the gauged supergravity. Still, the commutant (if any)  of $\,\textrm{G}\,$ inside $\,\mathcal{G}\,$ remains as a global symmetry of the theory after the gauging procedure.

In this work we will consider the case $\,n=6\,$. This is the largest value for which the duality group of half-maximal supergravity can be embedded into the one of maximal supergravity, \textit{i.e.} $\,\textrm{SL}(2) \times \textrm{SO}(6,6) \subset \textrm{E}_{7(7)}\,$, and the half-maximal supergravity Lagrangian can be embedded into the maximal one provided certain quadratic constraints on the couplings in the theory hold (see eq.(\ref{QC_Extra}) below). We investigate a specific class of gaugings of the group
\begin{equation}
\label{G_group}
\textrm{G}=\textrm{ISO}(3)_{1} \times \textrm{ISO}(3)_{2} \ , 
\end{equation}
which is embedded in the duality group as
\begin{equation}
\label{embedding_chain}
\textrm{G} \,\,\subset\,\, \textrm{SO}(3,3)_{1} \times \textrm{SO}(3,3)_{2}  \,\,\subset\,\, \textrm{SL}(2) \times \textrm{SO}(6,6) \ ,
\end{equation}
where we have attached labels $\,_1\,$ and $\,_2\,$ in order to keep track of each independent $\,\textrm{ISO}(3)\,$ and $\,\textrm{SO}(3,3)\,$ factor in (\ref{G_group}) and (\ref{embedding_chain}). General classes of gaugings of $\,\textrm{G}  \subset \textrm{SO}(3,3)_{1} \times \textrm{SO}(3,3)_{2}\,$ in half-maximal supergravity \cite{deRoo:2002jf,deRoo:2003rm,deRoo:2006ms} have been extensively investigated in the past, for example, with the aim of charting the landscape of flux compactifications \cite{Dibitetto:2010rg,Dibitetto:2011gm}.

In the duality-covariant formulation of \cite{Schon:2006kz}, the bosonic field content of the half-maximal supergravity consists of the metric $\,g_{\mu\nu}\,$, $\,12\,$ (electric) plus $\,12\,$ (magnetic) vector fields $\,A_\mu{}^{\alpha M}$, and $2+36$ scalar fields $\,\phi\textrm{'s}\,$ serving as coordinates in the coset space geometry
\begin{equation}
\label{M_scalar}
\mathcal{M}_{\textrm{scal}} = \frac{\textrm{SL}(2)}{\textrm{SO}(2)} \times \frac{\textrm{SO}(6,6)}{\textrm{SO}(6) \times \textrm{SO}(6)} \ ,
\end{equation}
and being parameterised by a coset representative $\,\mathcal{V}(\phi)\,$. Already at this level we have introduced a fundamental $\,\textrm{SL}(2)\,$ index $\,\alpha=\pm\,$ as well as a fundamental $\,\textrm{SO}(6,6)\,$ index $\,M$. These are raised/lowered using the $\,\epsilon_{\alpha\beta}\,$ and $\,\eta_{MN}\,$ invariant tensors of $\,\textrm{SL}(2)\,$ and $\,\textrm{SO}(6,6)$, respectively.

Having set the bosonic field content of the theory, all the interactions compatible with $\,{\mathcal{N}=4}\,$ supersymmetry and induced by gaugings of the type\footnote{The gauge group $\,\textrm{G}=\textrm{ISO}(3)_{1} \times \textrm{ISO}(3)_{2}\,$ is a particular example within the larger class of $\,\textrm{G}=\textrm{CSO}(p,q,r) \times \textrm{CSO}(p',q',r')\,$ gaugings investigated in \cite{deRoo:2006ms,Roest:2009tt}. The case of $\,\textrm{ISO}(3)\,$ corresponds to $\,{(p,q,r)=(3,0,1)}\,$ or $\,{(p,q,r)=(0,3,1)}\,$ (and equivalently for the primed factor).} (\ref{embedding_chain}) are encoded in a so-called \textit{embedding tensor}. This tensor comes along with an index structure $\,f_{\alpha MNP}=f_{\alpha [MNP]}\,$ and thus lives in the $\,(\textbf{2},\textbf{220})\,$ irreducible representation (irrep) of $\,\mathcal{G}\,$. It also specifies the non-Abelian gauge structure of the half-maximal gauged supergravity, namely,
\begin{equation}
\label{commutators_N=4}
[\, T_{\alpha M}\,,\,T_{\beta N}\,] = f_{\alpha MN}{}^{P} \, T_{\beta P} \ ,
\end{equation}
where $\,T_{\alpha M}\,$ are the generators of $\,\mathcal{G}\,$ that couple to the vector fields $\,A_{\mu}{}^{\alpha M}\,$ in the gauge connection. Consistency of the gauging procedure requires a set of quadratic constraints on the embedding tensor of the form \cite{Schon:2006kz}
\begin{equation}
\label{QC_N=4}
(\mathbf{3},\mathbf{495}): \,\,\, f_{\alpha [MN}{}^{R} \, f_{\beta PQ] R} = 0
\hspace{8mm} \textrm{ and } \hspace{8mm} 
(\mathbf{1},\mathbf{66}+\mathbf{2079}): \,\,\, \epsilon^{\alpha \beta} \, f_{\alpha MN}{}^{R} \, f_{\beta PQ R} = 0 \ ,
\end{equation}
where we have included the irrep of $\,\mathcal{G}\,$ where each constraint lives. In addition, there are two additional constraints given by
\begin{equation}
\label{QC_Extra}
(\mathbf{3},\mathbf{1}): \,\,\, f_{\alpha MNP} \, f_{\beta}{}^{MNP}= 0
\hspace{8mm} \textrm{ and } \hspace{8mm} 
(\mathbf{1},\mathbf{462'}): \,\,\, \left. \epsilon^{\alpha \beta} \, f_{\alpha [MNP} \, f_{\beta QRS]} \right|_{\textrm{SD}} = 0 \ ,
\end{equation}
where SD refers to the self-dual projection of the $\,\textrm{SO}(6,6)\,$ six-form. These two additional constraints (\ref{QC_Extra}) are \textit{not} required by $\,\mathcal{N}=4\,$ supersymmetry but must hold for the half-maximal Lagrangian to be embeddable into an $\,\mathcal{N}=8\,$ maximal gauged supergravity \cite{Dibitetto:2011eu}. Whenever these two additional constraints are satisfied, the half-maximal supergravity with gauging (\ref{G_group}) can be viewed as a subsector of the maximal gauged supergravity with gauging 
\begin{equation}
\textrm{G}_{\textrm{max}} = [\textrm{SO}(1,1) \times \textrm{SO}(6)] \ltimes \mathbb{R}^{12} \subset \textrm{E}_{7(7)} \ .    
\end{equation}
This maximal supergravity has recently received a lot of attention due to its connection to S-fold solutions in type IIB string theory as originally noticed in \cite{Inverso:2016eet}.

As a consequence of the gauging procedure, the fermions in the theory develop scalar-dependent mass terms and supersymmetry requires to introduce a non-trivial scalar potential. This is given by
\begin{equation}
\begin{array}{lll}
\label{V_N4}
V & = &  \frac{1}{64} \, f_{\alpha MNP} \, f_{\beta QRS} M^{\alpha \beta} \left[ \frac{1}{3} \, M^{MQ} \, M^{NR} \, M^{PS} + \left(\frac{2}{3} \, \eta^{MQ} - M^{MQ} \right) \eta^{NR} \eta^{PS} \right]  \\[3mm]
  & - & \frac{1}{144} \, \epsilon^{\alpha \beta}  \, f_{\alpha MNP} \, f_{\beta QRS} \, M^{MNPQRS}  \ ,
\end{array}
\end{equation}
where
\begin{equation}
\label{M_SL2}
M_{\alpha \beta} = \frac{1}{\textrm{Im}z_{7}}\left(  
\begin{array}{cc}
|z_{7}|^2   &  \textrm{Re}z_7  \\
\textrm{Re}z_7     &  1
\end{array}
\right) \in \textrm{SL}(2) \ ,
\end{equation}
encodes a complex scalar $\,z_{7}\,$ spanning the $\,\textrm{SL}(2)/\textrm{SO}(2)\,$ factor in (\ref{M_scalar}). Together with this, the potential depends on additional scalars spanning the $\,\textrm{SO}(6,6)/(\textrm{SO}(6)\times \textrm{SO}(6))\,$ factor in (\ref{M_scalar}). These are $\,36\,$ real scalars which can be assembled in a matrix
\begin{equation}
\label{M_SO66}
M_{MN} = \left(  
\begin{array}{cc}
G^{-1}   &  - G^{-1} \, B  \\
B \, G^{-1}    &  G - B \, G^{-1} \, B
\end{array}
\right)  \in \textrm{SO}(6,6) \ ,
\end{equation}
where $\,G\,$ and $\,B\,$ are arbitrary symmetric and anti-symmetric $\,6 \times 6\,$ matrices accounting for $\,21\,$ and $\,15\,$ scalars, respectively. For the class of gaugings in (\ref{G_group})-(\ref{embedding_chain}), the kinetic terms for the scalar fields serving as coordinates on the scalar geometry (\ref{M_scalar}) are constructed using standard coset techniques and read
\begin{equation}
\label{Lkin_general}
\mathcal{L}_{\textrm{kin}} =  \frac{1}{8} \, \partial_{\mu}M_{\alpha\beta} \, \partial^{\mu}M^{\alpha\beta}   +  \frac{1}{16} \, D_{\mu}M_{MN} \, D^{\mu}M^{MN} \ ,   
\end{equation}
where $\,M^{\alpha\beta}\,$ and $\,M^{MN}\,$ are the inverse of $\,M_{\alpha\beta}\,$ in (\ref{M_SL2}) and $\,M_{MN}\,$ in (\ref{M_SO66}), respectively. In this work we will focus on maximally symmetric AdS$_{4}$ solutions of the theory so that vector fields will be set to zero in the general covariant derivatives 
\begin{equation}
\label{cov_der_SO(6,6)}
D_{\mu}M_{MN} = \partial_{\mu}M_{MN} + 2 \, A_{\mu}{}^{\alpha P} \, f_{\alpha P(M}{}^{Q} \, M_{N)Q} \ .     
\end{equation}
Lastly, the scalar potential (\ref{V_N4}) depends on a specific $\,\textrm{SO}(6,6)\,$ six-form\footnote{Due to the $\,\epsilon_{mnpqrs}\,$ tensor with $\,{m,n,\ldots=1,\ldots,6}\,$ in the definition of the $\,\textrm{SO}(6,6)\,$ six-form (\ref{M_six-form}), the components $\,\mathcal{V}_{M}{}^{n}\,$ entering (\ref{M_six-form}) must be extracted from the coset representative $\,\mathcal{V}_{M}{}^{N}\,$ using a Lorentzian basis (for the column index $\,N\,$) of $\,\textrm{SO}(6,6)\,$ where $\,\eta_{MN}=\textrm{diag}(-\mathbb{I}_{6},\mathbb{I}_{6})\,$.}
\begin{equation}
\label{M_six-form}
M_{MNPQRS}  =   \epsilon_{mnpqrs} \, \mathcal{V}_{M}{}^{m} \, \mathcal{V}_{N}{}^{n} \, \mathcal{V}_{P}{}^{p} \, \mathcal{V}_{Q}{}^{q} \, \mathcal{V}_{R}{}^{r} \, \mathcal{V}_{S}{}^{s} \ ,
\end{equation}
that is constructed from the $\,\textrm{SO}(6,6)/(\textrm{SO}(6) \times \textrm{SO}(6))\,$ coset representative $\,\mathcal{V}_{M}{}^{N}\,$ such that $\,{M_{MN}=(\mathcal{V} \, \mathcal{V}^{t})_{MN}}\,$ (see \cite{Schon:2006kz} for more details).

The scalar potential (\ref{V_N4}) is a complicated function of the $\,2+36\,$ scalars spanning (\ref{M_scalar}). Trying to chart the landscape of $\,\mathcal{N}=2\,$ supersymmetric AdS$_{4}$ solutions by direct extremisation of (\ref{V_N4}) is out of computational reach, so we will resort to two simpler setups where only $\,2+12\,$ scalars are kept and the rest are set to zero. Following the original idea in \cite{Warner:1983vz}, this will be done in a group-theoretical consistent manner by retaining only those scalars that are invariant (singlets) under specific residual symmetry groups. In particular, we will consider discrete $\,\mathbb{Z}_{2}^2\,$ and continuous $\,\textrm{U}(1)_{\textrm{R}}\,$ subgroups of $\,\textrm{G}\,$ as such residual symmetry groups. In this manner, whenever an extremum is found in the simplified setup, it is guaranteed that it corresponds to an actual extremum of the scalar potential of the full theory. Moreover, although we will find extrema of the scalar potential in the setups with $\,2+12\,$ scalars, we will provide the full mass spectrum for all the bosonic and fermionic fields in half-maximal supergravity using \cite{Schon:2006kz,Borghese:2010ei}. This supergravity spectrum maps to the spectrum of operators in the would-be dual CFT$_3$'s.

\subsection{From $\,[\textrm{SO}(1,1) \times \textrm{SO}(6)] \ltimes \mathbb{R}^{12}\,$ to $\,\textrm{ISO}(3) \times \textrm{ISO}(3)\,$}

Starting from the $\,\textrm{G}_{\textrm{max}}=[\textrm{SO}(1,1) \times \textrm{SO}(6)] \ltimes \mathbb{R}^{12}\,$ gauged maximal supergravity of \cite{Inverso:2016eet}, and modding it out by a discrete subgroup $\,\mathbb{Z}_{2} \subset \textrm{G}_{\textrm{max}}\,$ \cite{Dibitetto:2011eu}, one is left with a very specific gauging
\begin{equation}
\label{G_group_sec_2.1}
\textrm{G}=\textrm{ISO}(3)_{1} \times \textrm{ISO}(3)_{2} \ ,    
\end{equation}
of half-maximal supergravity. In order to describe the resulting half-maximal supergravity, it will prove convenient to first perform a light-cone splitting $\,M=(m,\bar{m})\,$ with respect to the $\,\textrm{SO}(6,6)\,$ invariant metric
\begin{equation}
\eta_{MN} = \left( \begin{array}{cc}
0 & \delta_{m \bar{n}} \\
\delta_{\bar{m}n} & 0
\end{array} \right) 
\hspace{8mm} \textrm{ with } \hspace{8mm} 
m=1,\ldots,6  \hspace{4mm} \textrm{ , } \hspace{4mm}  \bar{n}=\bar{1},\ldots,\bar{6} \ ,
\end{equation}
and then a further splitting $\,m=(a,i)\,$ and $\,\bar{m}=(\bar{a},\bar{i})\,$ with $\,a = 1,3,5\,$ and $\,i = 2,4,6\,$. In this manner, the original $\,\textrm{SO}(6,6)\,$ fundamental index $\,M\,$ has a decomposition
\begin{equation}
\begin{array}{ccccccc}
\textrm{SO}(6,6) & \supset & \textrm{SO}(3,3)_{1} &\times& \textrm{SO}(3,3)_{2} \\[2mm]
    M  & \rightarrow & (\, a \,,\, \bar{a}\, ) &\oplus&  ( \, i \,,\, \bar{i}\,)
\end{array}
\end{equation}
and the $\,\textrm{ISO}(3)_{1,2}\,$ factors in (\ref{G_group_sec_2.1}) are embedded into $\,\textrm{SO}(3,3)_{1,2} \sim \textrm{SL}(4)_{1,2}\,$, respectively. An explicit computation of the resulting embedding tensor $\,f_{\alpha MNP}\,$ specifying the half-maximal supergravity yields (using conventions in \cite{Dibitetto:2011gm})
\begin{equation}
\label{ET_N=8}
\begin{array}{lclclc}
f_{+\bar{a}bc} = 2 \, g  & \hspace{5mm}   ,   &  \hspace{5mm}  f_{-abc} = \pm \, 2  \, g \, c & , \\[2mm]
f_{-i\bar{j}\bar{k}} = 2  \, g  & \hspace{5mm}   ,   &  \hspace{5mm}  f_{+\bar{i}\bar{j}\bar{k}} = \pm \, 2 \, g \, c   & ,
\end{array}
\end{equation}
where $\,g\,$ is the gauge coupling and $\,c\,$ is a parameter encoding the dyonic nature of the gauging, and with all the other components vanishing. In what follows we are assuming a cyclic structure in all the embedding tensor components of the same type, \textit{i.e.} $\,f_{+ \bar{1}35} = f_{+ \bar{3}51} = f_{+ \bar{5}13} = 2 \, g\,$, etc., so we are intentionally omitting epsilon symbols in (\ref{ET_N=8}) to lighten the notation. Lastly, since the embedding tensor in (\ref{ET_N=8}) is the result of \textit{halving} the $\,\textrm{G}_{\textrm{max}}=[\textrm{SO}(1,1) \times \textrm{SO}(6)] \ltimes \mathbb{R}^{12}\,$ gauging of maximal supergravity, it automatically satisfies the extra constraints in (\ref{QC_Extra}) for a half-maximal supergravity to be embeddable in a maximal supergravity.

\subsection{Deforming $\,\textrm{ISO}(3) \times \textrm{ISO}(3)\,$ half-maximal supergravity}
\label{sec:Deforming_ISO(3)xISO(3)}

Following \cite{Roest:2009tt} (in the conventions of \cite{Dibitetto:2011gm}), we will deform the $\,\textrm{G}=\textrm{ISO}(3)_{1} \times \textrm{ISO}(3)_{2}\,$ gauging specified by (\ref{ET_N=8}) while preserving the $\,\mathcal{N}=4\,$ supersymmetry of half-maximal supergravity. We are doing so by activating two additional embedding tensor components
\begin{equation}
f_{+abc}
\hspace{8mm} \textrm{ and } \hspace{8mm} 
f_{-\bar{i}\bar{j}\bar{k}} \ .
\end{equation}
As discussed in \cite{Roest:2009tt,Dibitetto:2011gm}, turning on these two components modifies how the gauge group $\,{\textrm{G}=\textrm{ISO}(3)_{1} \times \textrm{ISO}(3)_{2}}\,$ is embedded into the duality group, see (\ref{embedding_chain}). We will change this embedding in a parametrically controlled manner yielding a one-parameter generalisation of the gauging in (\ref{ET_N=8}).

Let us denote $\,\tilde{\varphi}\,$ the new parameter entering the embedding tensor, which now has components
\begin{equation}
\label{ET_N=4_N=3_vac}
\begin{array}{lclclc}
f_{+\bar{a}bc} = \frac{ 2 \, \sqrt{2} \, g}{\sqrt{1 + \tilde{\varphi}^2 }}  &  \hspace{2mm}  ,   &  \hspace{2mm}  f_{-abc}=  \pm 2 \sqrt{2} \, g \, c \, \frac{\tilde{\varphi}}{\sqrt{1 + \tilde{\varphi}^2 }} & \hspace{2mm}   ,   &  \hspace{2mm}  f_{+abc}  = - 2 \sqrt{2}\, g \, c \,  \frac{\tilde{\varphi}^2-1}{\tilde{\varphi}^2+1}  & , \\[4mm]
f_{-i\bar{j}\bar{k}}  = \frac{2 \, \sqrt{2} \, g}{\sqrt{1 + \tilde{\varphi}^2 }}   & \hspace{2mm}   ,   &\hspace{2mm}  f_{+\bar{i}\bar{j}\bar{k}}= \pm  2 \sqrt{2} \, g \, c \, \frac{\tilde{\varphi}}{\sqrt{1 + \tilde{\varphi}^2 }}    &   \hspace{2mm} ,   &  \hspace{2mm} f_{-\bar{i}\bar{j}\bar{k}} =  - 2 \sqrt{2}\, g \, c \,  \frac{\tilde{\varphi}^2-1}{\tilde{\varphi}^2+1} & ,
\end{array}
\end{equation}
and, as we will see in a moment, accommodates a rich structure of new AdS$_{4}$ vacua. The class of gaugings in (\ref{ET_N=4_N=3_vac}) automatically solves the quadratic constrains in (\ref{QC_N=4}) required by half-maximal supersymmetry. However, an explicit computation of the additional quadratic constraints in (\ref{QC_Extra}) yields
\begin{equation}
\label{QC_Extra_N=3_vac}
f_{\alpha MNP} \, f_{\beta}{}^{MNP}= 0
\hspace{10mm} \textrm{ and } \hspace{10mm}
\left. \epsilon^{\alpha \beta} \, f_{\alpha [MNP} \, f_{\beta QRS]} \right|_{\textrm{SD}}  \propto  g^2 c \,  \dfrac{\tilde{\varphi}^2-1}{\left(\tilde{\varphi}^2+1\right)^{\frac{3}{2}}} \ .
\end{equation}
As a result, due to the violation of the constraint living in the $\,(\mathbf{1},\mathbf{462'})\,$ irrep, the deformed theories do not admit an uplift to maximal supergravity unless $\,\tilde{\varphi}^2 = 1\,$. Note that, at $\,\tilde{\varphi}^2 = 1\,$, the embedding tensor in (\ref{ET_N=4_N=3_vac}) consistently reduces to the one in (\ref{ET_N=8}) and the theory becomes a $\,\mathbb{Z}_{2}$-invariant subsector of the $\,\textrm{G}_{\textrm{max}}=[\textrm{SO}(1,1) \times \textrm{SO}(6)] \ltimes \mathbb{R}^{12}\,$ maximal supergravity.

\subsection{The gauge algebra of the deformed $\,\mathcal{N}=4\,$ models}
\label{sec:algebra_ISO(3)xISO(3)}

It is instructive to take a closer look at how the precise $\,\mathcal{N}=4\,$ gauging specified by (\ref{ET_N=4_N=3_vac}) is realised at an algebraic level in terms of the generators
\begin{equation}
T_{\alpha M} = \left( \, T_{\alpha a} \, , \,  T_{\alpha \bar{a}} \,\,\, ; \,\,\, T_{\alpha i} \, , \, T_{\alpha \bar{i}} \, \right)
\end{equation}
entering the commutation relations (\ref{commutators_N=4}). We will show that each of the $\,\textrm{ISO}(3)_{1,2}\,$ factors is at a different $\,\textrm{SL}(2)\,$ angle in the spirit of \cite{deRoo:1985jh}. Moreover, each $\,\textrm{ISO}(3)_{1,2}\,$ factor by itself involves a non-trivial $\,\textrm{SO}(3,3)_{1,2}\,$ angle (both unbar/bar generators are present) in the spirit of  \cite{Roest:2009tt}.

Before discussing its gauge structure in detail, it is also worth noticing that the class of half-maximal $\,\textrm{G}=\textrm{ISO}(3)_{1} \times \textrm{ISO}(3)_{2}\,$ gaugings specified by the embedding tensor (\ref{ET_N=4_N=3_vac}) depends on three arbitrary parameters $\,(g,c,\tilde{\varphi})\,$. However the most general class of gaugings of $\,\textrm{G}=\textrm{ISO}(3)_{1} \times \textrm{ISO}(3)_{2}\,$ in half-maximal supergravity involves eight parameters (up to gauge fixing) and is discussed in detail in Appendix~\ref{app:ISO(3)xISO(3)_general_gauging}. The study of the structure of AdS$_{4}$ solutions in this more general class of models goes beyond the scope of this work and is postponed for the future.

The algebraic realisation of the $\,\textrm{G}=\textrm{ISO}(3)_{1} \times \textrm{ISO}(3)_{2}\,$ gaugings specified by (\ref{ET_N=4_N=3_vac}) involves a set of $\,12\,$ independent generators. For example, choosing them to be $\,(T_{+ a},T_{+ \bar{a}})\,$ and $\,(T_{- i},T_{- \bar{i}})\,$, the antisymmetry of the brackets (\ref{commutators_N=4}) further sets
\begin{equation}
 T_{-a}  = c \, \tilde{\varphi}  \,\, T_{+ \bar{a}}
\hspace{5mm} , \hspace{5mm}
T_{+ \bar{i}}  = c \, \tilde{\varphi} \,\, T_{- i}
\hspace{5mm} , \hspace{5mm}
T_{- \bar{a}} = T_{+ i } = 0 \ ,
\end{equation}
and the independent generators satisfy non-trivial commutation relations of the form
\begin{equation}
\label{algebra_ISO3_1_family_I}
\begin{array}{llll}
\left[ \,  T_{+ a} , T_{+ b} \, \right] &=& 2 \sqrt{2}\, g \, \left(   \dfrac{1}{\sqrt{1+\tilde{\varphi}^2}} \, \epsilon_{ab}{}^{c} \,  T_{+ c}  + c \,  \dfrac{1-\tilde{\varphi}^2}{1+\tilde{\varphi}^2} \,    \epsilon_{ab}{}^{\bar{c}} \,  \, T_{+ \bar{c}} \right) \ , \\[4mm]
\left[ \, T_{+ a} , T_{+ \bar{b}} \, \right] &=& 2 \sqrt{2}\, g \, \dfrac{1}{\sqrt{1+\tilde{\varphi}^2}} \, \epsilon_{a\bar{b}}{}^{\bar{c}} \,  T_{+ \bar{c}} \ , \\[4mm]
\left[ \, T_{+ \bar{a}} , T_{+ \bar{b}} \, \right] &=& 0 \ ,
\end{array}
\end{equation}
for the $\,\textrm{ISO}(3)_{1}\,$ factor in the gauge group and, similarly,
\begin{equation}
\label{algebra_ISO3_2_family_I}
\begin{array}{llll}
\left[ \,  T_{- \bar{i}} , T_{- \bar{j}} \, \right] &=&  2 \sqrt{2}\, g \,  \left(  \dfrac{1}{\sqrt{1+\tilde{\varphi}^2}}  \, \epsilon_{\bar{i}\bar{j}}{}^{\bar{k}} \, T_{- \bar{k}}   + \, c \,  \dfrac{1-\tilde{\varphi}^2}{1+\tilde{\varphi}^2}  \, \epsilon_{\bar{i}\bar{j}}{}^{k} \, T_{- k} \right) \ , \\[4mm]
\left[ \, T_{- \bar{i}} , T_{- j} \, \right] &=& 2 \sqrt{2}\, g \, \dfrac{1}{\sqrt{1+\tilde{\varphi}^2}} \, \epsilon_{\bar{i} j}{}^{k}  \,  T_{- k} \ , \\[4mm]
\left[ \, T_{- i}  ,  T_{- j} \, \right] &=& 0 \ ,
\end{array}
\end{equation}
for the $\,\textrm{ISO}(3)_{2}\,$ factor. There is the limiting case $\,\tilde{\varphi} \rightarrow \pm \infty\,$ for which the embedding tensor (\ref{ET_N=4_N=3_vac}) stays regular and the $\,\textrm{ISO}(3)_{1,2}\,$ factors become nilpotent. More concretely, they reduce to the nilpotent algebra denoted $\,n(3.5)\,$ in Table~$4$ of \cite{Grana:2006kf}. The drastic change in the four-dimensional algebra structure at $\,\tilde{\varphi} \rightarrow \pm \infty\,$ suggests a drastic change in the interpretation of the corresponding supergravity solutions as well as of their possible uplifts to ten or eleven dimensions.

\section{$\mathbb{Z}_{2}^2$-invariant sector}
\label{sec:Z2xZ2_sector}

We will look at the $\,\mathbb{Z}_{2}^2$-invariant sector of half-maximal supergravity. Being an invariant sector with respect to a (in this case discrete) $\,\mathbb{Z}_{2}^2\,$ subgroup of $\,\textrm{G}\,$, an extremum of the scalar potential in this simplified setup automatically implies an extremum of the full scalar potential.

\subsection{The $\,\mathcal{N}=1\,$ seven-chiral model}

The $\,\mathbb{Z}_{2}^2$-invariant sector of half-maximal supergravity was investigated in \cite{Aldazabal:2008zza}. It can be recast as a minimal $\,\mathcal{N}=1\,$ supergravity coupled to seven chiral multiplets. We will denote $\,z_{I}\,$, with $\,I=1,\ldots,7\,$, the seven complex scalars in the chiral multiplets. One of them, we choose it to be $\,z_{7}\,$, is the one parameterising the $\,M_{\alpha \beta} \in \textrm{SL}(2)\,$ element in (\ref{M_SL2}). The remaining six complex fields specify the $\,G\,$ and $\,B\,$ matrices in (\ref{M_SO66}) from which the $\,M_{MN} \in \textrm{SO}(6,6)\,$ element is constructed. More concretely,
\begin{equation}
G = \left(  
\begin{array}{ccc}
G_1   &  0 & 0 \\
0  & G_2 & 0 \\
0 &  0 & G_3 \\
\end{array}
\right)
\hspace{8mm} \textrm{ and } \hspace{8mm}
B = \left(  
\begin{array}{ccc}
B_1   &  0 & 0 \\
0  & B_2 & 0 \\
0 &  0 & B_3 \\
\end{array}
\right)
\end{equation}
are block-diagonal matrices with components
\begin{equation}
\label{G&B_blocks}
G_{i} = \frac{\textrm{Im}z_{i+3}}{\textrm{Im}z_{i}}\left(  
\begin{array}{cc}
1  &  \textrm{Re}z_i  \\
\textrm{Re}z_i     &  |z_{i}|^2 
\end{array}
\right)
\hspace{4mm} , \hspace{4mm}
B_{i} = \left(  
\begin{array}{cc}
0  &  \textrm{Re}z_{i+3}  \\
-\textrm{Re}z_{i+3}     & 0
\end{array}
\right)
\hspace{4mm} , \hspace{4mm}
i=1,2,3 \ ,
\end{equation}
that depend on the complex scalars $\,z_{1}, \ldots, z_{6}\,$. The scalar kinetic terms for this sector of the theory take the form
\begin{equation}
\mathcal{L}_{\textrm{kin}} = - \frac{1}{4} \, \sum_{I=1}^{7} \left( \, d\varphi_{I}^2  + e^{2 \varphi_{I}}  \, d\chi_{7}^2 \, \right)
\hspace{8mm} \textrm{ with } \hspace{8mm}
z_{I}= -\chi_{I} + i \, e^{-\varphi_{I}} \ .
\end{equation}
The scalar manifold invariant under the $\,\mathbb{Z}_{2}^{2}\,$ discrete symmetry is therefore identified with the special K\"ahler (SK) factorised geometry
\begin{equation}
\mathcal{M}^{^{\mathbb{Z}_{2}^{2}}}_{\textrm{scal}} = \left[ \frac{\textrm{SL}(2)}{\textrm{SO}(2)} \right]^{7} \subset \, \frac{\textrm{SL}(2)}{\textrm{SO}(2)} \times \frac{\textrm{SO}(6,6)}{\textrm{SO}(6) \times \textrm{SO}(6)} \ .
\end{equation}

In the $\,\mathbb{Z}_{2}^2$-invariant sector of half-maximal supergravity, the scalar potential takes a lengthy but more tractable expression in terms of the seven complex scalars $\,z_{I}\,$. We will find families of $\,\mathcal{N}=2\,$ supersymmetric AdS$_{4}$ extrema analytically in this setup.

\subsection{Warming up: the $\varphi$-family of $\,\mathcal{N}=2\,$ AdS$_{4}$ solutions of \cite{Bobev:2021yya}}

The half-maximal gauged supergravity specified by the undeformed ($\tilde{\varphi}=\pm 1$) embedding tensor (\ref{ET_N=8}) possesses a one-parameter $\varphi$-family of $\,\mathcal{N}=2\,$ supersymmetric AdS$_{4}$ solutions.\footnote{These solutions were originally constructed in \cite{Bobev:2021yya} within the context of $\,\textrm{G}_{\textrm{max}}=[\textrm{SO}(1,1) \times \textrm{SO}(6)] \ltimes \mathbb{R}^{12}\,$ maximal gauged supergravity. Two additional moduli fields $\,\chi_{1,2}\,$ dual to marginal operators and dubbed axion-like flat deformations in \cite{Guarino:2021hrc} were also identified for this family of AdS$_{4}$ solutions. In an appropriate basis, the combination $\,\chi \equiv \chi_{1}-\chi_{2}\,$ preserves $\,\mathcal{N}=2\,$ supersymmetry whereas the orthogonal combination breaks supersymmetry completely. We will come back to the axion-like flat deformation $\,\chi\,$ in Section~\ref{sec:U(1)_sector} when exploring a $\,\textrm{U}(1)_{\textrm{R}}$-invariant sector of the theory.} This $\varphi$-family of solutions lies at the loci
\begin{equation}
\label{vacuum_N=8_N=2_vac}
z_{1} = - \bar{z}_{3}= i \, c \, \sqrt{\frac{1+\varphi^2}{2}}
\hspace{2mm} , \hspace{2mm}
z_{2} = i \, c
\hspace{2mm} , \hspace{2mm}
z_{4}=-\bar{z}_{6}= \dfrac{-\varphi + i}{\sqrt{1+\varphi^2}} 
\hspace{2mm} , \hspace{2mm}
z_{5}= z_{7}= \dfrac{\mp 1 + i}{\sqrt{2}}  \ ,
\end{equation}
where the $\,\mp\,$ sign in (\ref{vacuum_N=8_N=2_vac}) is correlated with the $\,\pm\,$ sign in (\ref{ET_N=8}). Having $\,\textrm{Im}z_{1,2,3}>0\,$ then requires $\,c>0\,$. The vacuum energy turns out to be independent of $\,\varphi\,$ and given by 
\begin{equation}
\label{V0_N8}
V_{0}=-3 \, g^2 \, c^{-1} \ .
\end{equation}

The AdS$_{4}$ solutions in (\ref{vacuum_N=8_N=2_vac}) preserve $\,\mathcal{N}=2\,$ supersymmetry and a $\,\textrm{U}(1)_{\textrm{R}}\,$ residual symmetry at generic values of $\,\varphi\,$ within half-maximal supergravity\footnote{$\mathcal{N}=2$ supersymmetry and $\,\textrm{U}(1)_{\textrm{R}}\times \textrm{U}(1)_{\textrm{F}}\,$ symmetry in the maximal theory without the $\mathbb{Z}_{2}$ projection. To set up notation, we have attached labels $_{\textrm{R}}$ and $_{\textrm{F}}$ to identify the corresponding R-symmetry and flavour symmetry groups in the would-be dual SCFT$_{3}$'s}. However, and even though the vacuum energy in (\ref{V0_N8}) does not depend on $\,\varphi\,$, there are two special values of $\,\varphi\,$ at which (super) symmetry enhancements occur:

\begin{itemize}

\item \textbf{Point $\,\varphi=0\,$:} At this value the AdS$_{4}$ solution preserves $\,\mathcal{N}=2\,$ supersymmetry and a $\,\textrm{U}(1)_{\textrm{R}}\times \textrm{U}(1)_{\textrm{F}}\,$ symmetry within half-maximal supergravity.\footnote{$\mathcal{N}=2$ supersymmetry and $\,\textrm{U}(1)_{\textrm{R}}\times\textrm{SU}(2)_{\textrm{F}}\,$ symmetry in the maximal theory without the $\mathbb{Z}_{2}$ projection.}\\[-3mm]

\item \textbf{Point $\,\varphi=\pm 1\,$:} At these values the AdS$_{4}$ solution preserves $\,\mathcal{N}=3\,$ supersymmetry and an $\,\textrm{SO}(3)_{\textrm{R}}\,$ symmetry within the half-maximal supergravity.\footnote{$\mathcal{N}=4$ supersymmetry and $\,\textrm{SO}(4)_{\textrm{R}}\,$ symmetry in the maximal theory without the $\mathbb{Z}_{2}$ projection.}

\end{itemize}

\noindent The AdS$_{4}$ solutions at $\,\varphi=0\,$ and $\,\varphi =  \pm 1\,$  uplift to the $\,\mathcal{N}=2\,$ and $\,\mathcal{N}=4\,$ type IIB S-folds constructed in \cite{Guarino:2020gfe} and \cite{Inverso:2016eet}, respectively. From a four-dimensional perspective, the parameter $\,\varphi\,$ appears to be non-compact. From a higher-dimensional perspective, the compactness of $\,\varphi\,$ remains an open question since the type IIB uplift of the entire $\,\varphi$-family of AdS$_4$ solutions in (\ref{vacuum_N=8_N=2_vac}) has not been constructed yet.

\subsubsection*{Marginal deformation and $\mathfrak{osp}(2|4)\,$ superconformal multiplets}

Being a flat direction in the scalar potential, $\,\varphi\,$ was identified with a marginal deformation specifying a direction in an $\,\mathcal{N}=2\,$ conformal manifold of S-fold CFT$_{3}$'s \cite{Bobev:2021yya}.\footnote{The apparent non-compactness of $\,\varphi\,$ (see \cite{Cesaro:2021tna} for a KK approach to this question) poses some challenges to the CFT Distance Conjecture \cite{Perlmutter:2020buo}.} Interestingly, there are unprotected operators in the $\,\mathcal{N}=2\,$ S-fold CFT$_{3}$'s whose conformal dimensions depend on $\,\varphi\,$.

According to the AdS$_{4}$/CFT$_{3}$ correspondence, the mass spectrum of the full set of half-maximal supergravity fields at the $\,\mathcal{N}=2\,$ AdS$_{4}$ solutions in (\ref{vacuum_N=8_N=2_vac}) can be arranged into multiplets of the $\,\mathfrak{osp}(2|4)\,$ superconformal symmetry of the dual $\,\mathcal{N}=2\,$ CFT$_{3}$'s. Following the notation\footnote{\label{conventions_N=2}For the $\,\mathcal{N}=2\,$ supermultiplets in three dimensions, our conventions for the Lorentz and R-symmetry Dynkin labels differ from the one in \cite{Cordova:2016emh}: $\,j=\frac{1}{2} j_{\tiny{\cite{Cordova:2016emh}}}\,$ and $\,R=r_{\tiny{\cite{Cordova:2016emh}}}\,$.} of \cite{Cordova:2016emh} for a superconformal multiplet $\,[j]_{\Delta}^{R}\,$, where $\,j\,$ and $\,R\,$ are the Lorentz and R-symmetry Dynkin labels of the highest weight state (HWS) in the multiplet and $\,\Delta\,$ is its conformal dimension, the spectrum contains five unprotected long multiplets
\begin{equation}
\label{long_maximal_theory}
L\bar{L}[0]^0_{\Delta_1} 
\hspace{5mm} ,  \hspace{5mm} 
L\bar{L}[\tfrac{1}{2}]^{0}_{\Delta_\pm}
\hspace{5mm} ,  \hspace{5mm} 
L\bar{L}[0]^{0}_{\tilde{\Delta}_{\pm}}  \ ,
\end{equation}
with conformal dimensions given by
\begin{equation}
\label{Delta_maximal_theory}
\begin{array}{rcl c rcll}
\Delta_1 &=& \tfrac{1}{2} + \tfrac{1}{2} \sqrt{\tfrac{17 + 33\varphi^2}{1 + \varphi^2}}
& \hspace{5mm} , & \hspace{5mm}
\Delta_\pm &=& \tfrac{1}{2} + \tfrac{2+(\varphi \pm 1)\varphi}{\sqrt{2(1+\varphi^2)}} & ,\\[4mm]
\tilde{\Delta}_{-}  &=& \tfrac{1}{2} + \tfrac{1}{2} \sqrt{1 + 8 \varphi^2}
& \hspace{5mm} , & \hspace{5mm}
\tilde{\Delta}_{+} &=& \tfrac{1}{2} + \tfrac{1}{2} \sqrt{\tfrac{17 + \varphi^2}{1+\varphi^2}}  & . 
\end{array}
\end{equation}
In addition there are one short and two semi-short protected multiplets with integer conformal dimension $\,\Delta=2\,$, namely,
\begin{equation}
\label{short_maximal_theory}
A_1\bar{A}_1[1]_2^0
\hspace{5mm} , \hspace{5mm}
L\bar{B}_1[0]_2^2
\hspace{5mm} , \hspace{5mm}
B_1\bar{L}[0]_2^{-2} \ ,
\end{equation}
where $\,A_1\bar{A}_1[1]^{0}_2\,$ is the stress-energy tensor multiplet of the $\,\mathcal{N}=2\,$ CFT$_{3}$'s.

The multiplets in (\ref{long_maximal_theory})-(\ref{short_maximal_theory}) describe a $\,\mathbb{Z}_{2}$-invariant subset of the spectrometry performed in \cite{Bobev:2021yya} within the context of the $\,\textrm{G}_{\textrm{max}}=[\textrm{SO}(1,1) \times \textrm{SO}(6)] \ltimes \mathbb{R}^{12}\,$ maximal supergravity.  The two semi-short multiplets $\,L\bar{B}_1[0]_2^2\,$ and $\,B_1\bar{L}[0]_2^{-2}\,$ in (\ref{short_maximal_theory}) contain the two real marginal operators investigated in \cite{Bobev:2021yya}. They are $\,\mathbb{Z}_{2}$-even enabling us to capture them also within the context of half-maximal supergravity. The scalar modulus $\,\varphi\,$ in (\ref{vacuum_N=4_N=2}) is dual to one such marginal operators. The other marginal operator is dual to a different modulus $\,\chi\,$ that will be studied in detail in Section~\ref{sec:U(1)_sector}. Last but not least, the spectrum in (\ref{long_maximal_theory})-(\ref{short_maximal_theory}) does \textit{not} contain a $\,\mathfrak{u}(1)_\textrm{F}\,$ flavour current multiplet $\,A_2\bar{A}_2[0]_1^0\,$ present in \cite{Bobev:2021yya}. This multiplet is $\,\mathbb{Z}_{2}$-odd and therefore projected away when truncating from maximal to half-maximal supergravity. This fact has some consequences we touch upon in the conclusions.

\subsection{A $\,(\varphi \,;\tilde{\varphi})$-family of $\,\mathcal{N}=2\,$  AdS$_{4}$ solutions}
\label{sec:family_1}

Let us now consider the effect of turning on the deformation parameter $\,\tilde{\varphi}\,$, \textit{i.e.}  $\,\tilde{\varphi} \neq \pm 1\,$ in the embedding tensor (\ref{ET_N=4_N=3_vac}). As already anticipated, turning on this parameter produces new $\,\mathcal{N}=2\,$ supersymmetric AdS$_{4}$ solutions which can still be found analytically.

At generic values of the deformation parameter $\,\tilde{\varphi}\,$, the locus of the $\,\mathcal{N}=2\,$ AdS$_{4}$ solutions in (\ref{vacuum_N=8_N=2_vac}) changes to 
\begin{equation}
\label{vacuum_N=4_N=2}
z_{1} = - \bar{z}_{3}= i \, c \, \sqrt{\dfrac{1+\varphi^2}{1+\tilde{\varphi}^2}}
\hspace{2mm} , \hspace{2mm}
z_{2} = i \, c
\hspace{2mm} , \hspace{2mm}
z_{4}=-\bar{z}_{6}= \dfrac{-\varphi + i}{\sqrt{1+\varphi^2}} 
\hspace{2mm} , \hspace{2mm}
z_{5}= z_{7}= \dfrac{\mp\tilde{\varphi} + i}{\sqrt{1+\tilde{\varphi}^2}}  \ ,
\end{equation}
with the $\,\mp\,$ sign in (\ref{vacuum_N=4_N=2}) being correlated with the $\,\pm\,$ sign in (\ref{ET_N=4_N=3_vac}). Notice that having $\,\textrm{Im}z_{1,2,3}>0\,$ still requires $\,c>0\,$, and also the reflection symmetry $\,\tilde{\varphi} \rightarrow -\tilde{\varphi}\,$. The vacuum energy turns out to be independent of the embedding tensor deformation $\,\tilde{\varphi}\,$ and, therefore, still given by 
\begin{equation}
\label{V0_Class_I}
V_{0}= - 3 \, g^2 \, c^{-1}  \ .
\end{equation}
Taking the limit $\,\tilde{\varphi} \rightarrow \pm \infty\,$ becomes pathological as $\,\textrm{Im}z_{1,3,5,7}=0\,$ hinting at some decompactification regime. This decompactification regime resonates well with the fact that taking $\,\tilde{\varphi} \rightarrow \pm \infty\,$ changes the gauge group to a new one being nilpotent (see discussion below (\ref{algebra_ISO3_1_family_I})-(\ref{algebra_ISO3_2_family_I})).

\subsection{$\mathfrak{osp}(2|4)\,$ superconformal multiplets}

The full half-maximal supergravity spectrum at this $\,(\varphi \,;\tilde{\varphi})$-family of AdS$_{4}$ solutions can be arranged into multiplets of the $\,\mathfrak{osp}(2|4)\,$ superconformal symmetry of the dual $\,\mathcal{N}=2\,$ CFT$_{3}$'s. The spectrum contains five unprotected long multiplets
\begin{equation}
\label{long_multiplets_general_Z2xZ2}
L\bar{L}[0]^0_{\Delta_1} 
\hspace{5mm} ,  \hspace{5mm} 
L\bar{L}[\tfrac{1}{2}]^{0}_{\Delta_\pm}
\hspace{5mm} ,  \hspace{5mm} 
L\bar{L}[0]^{0}_{\tilde{\Delta}_\pm}  \ ,
\end{equation}
with conformal dimensions given by
\begin{equation}
\label{Deltas_general_Z2xZ2}
\begin{array}{lcl}
\Delta_1 &\hspace{-2mm}=& \hspace{-2mm} \frac{1}{2}  + \frac{1}{2}   \sqrt{\frac{9+25 \tilde{\varphi}^2+\varphi ^2 \left(17+49 \tilde{\varphi}^2\right)}{\left(1+\varphi^2\right) \left(1+\tilde{\varphi}^2\right)}} \ , \\[6mm]
\Delta_\pm &\hspace{-2mm}=& \hspace{-2mm} 1 + \frac{1}{2}\sqrt{\frac{4 \varphi ^4+\varphi ^2 \left(13 \tilde{\varphi}^2+9\right)+4 \tilde{\varphi}^4+9 \tilde{\varphi}^2+5}{\left(\varphi ^2+1\right) \left(\tilde{\varphi}^2+1\right)}-\frac{4 \left(\varphi ^2+\tilde{\varphi}^2+1\right)\pm 8 \varphi ^3 \tilde{\varphi}\pm4 \varphi  \tilde{\varphi} \left(2 \tilde{\varphi}^2+2-\sqrt{\left(\varphi ^2+1\right) \left(\tilde{\varphi}^2+1\right)}\right)}{\sqrt{\left(\varphi ^2+1\right) \left(\tilde{\varphi}^2+1\right)}}} \ ,\\[6mm]
\tilde{\Delta}_\pm  &\hspace{-2mm}=& \hspace{-2mm} \frac{1}{2}+\frac{1}{2}\sqrt{\frac{5 \left( 1 + \tilde{\varphi}^2\right) +8 \left( \varphi^4+\tilde{\varphi}^4\right)+\varphi ^2 \left(\tilde{\varphi}^2+9\right)\pm 4 \sqrt{4 \varphi ^8+8 \varphi ^6-4 \varphi ^4 \left(\tilde{\varphi}^4+3 \tilde{\varphi}^2-1\right)-4 \varphi ^2 \tilde{\varphi}^2 \left(3+\tilde{\varphi}^2\right)+\left(1+\tilde{\varphi}^2+2 \tilde{\varphi}^4\right)^2}}{\left(1+\varphi^2\right) \left(1+\tilde{\varphi}^2\right)}}  \ .
\end{array}
\end{equation}
In addition, there are one short and two semi-short protected multiplets with integer conformal dimension $\,\Delta=2\,$. These are the same as in (\ref{short_maximal_theory}), namely,
\begin{equation}
\label{short&semi-short_multiplets_general_Z2xZ2}
A_1\bar{A}_1[1]^{0}_{2} 
\hspace{5mm} ,  \hspace{5mm} 
L\bar{B}_1[0]^{2}_{2} 
\hspace{5mm} ,  \hspace{5mm} 
B_1\bar{L}[0]^{-2}_{2}   \ ,
\end{equation}
where $\,A_1\bar{A}_1[1]^{0}_2\,$ is the stress-energy tensor multiplet of the dual $\,\mathcal{N}=2\,$ CFT$_{3}$'s.

\subsection{Special loci}

The $\,(\varphi \,;\tilde{\varphi})$-family of $\,\mathcal{N}=2\,$ AdS$_{4}$ solutions in (\ref{vacuum_N=4_N=2}) generically preserves a $\,\textrm{U}(1)_{\textrm{R}}\,$ symmetry. The latter can be seen from the normalised gravitino masses which are given by
\begin{equation}
\label{Gravitino_masses_N4}
\begin{array}{lll}
m L  &=&  1 \,\,\,\, (\times 2) 
\hspace{5mm} , \hspace{5mm}
\dfrac{1 + \varphi^2  + \tilde{\varphi}^2  \pm \varphi \, \tilde{\varphi}}{\sqrt{\left( 1+\varphi^2 \right) \left( 1+\tilde{\varphi}^2 \right)}}  \ .
\end{array}
\end{equation}
As a result, the marginal deformation $\,\varphi\,$ turns out to be compatible with the embedding tensor deformation $\,\tilde{\varphi}\,$. Moreover, they both enter the AdS$_{4}$ solutions (\ref{vacuum_N=4_N=2}) and the normalised gravitino masses (\ref{Gravitino_masses_N4}) in a very symmetric fashion.

A detailed inspection of the normalised gravitino masses in (\ref{Gravitino_masses_N4}) singles out four special cases to be further investigated:
\begin{equation}
\label{special_cases}
\begin{array}{lllll}
i) & \varphi = \tilde{\varphi} = 0 &  \Rightarrow &  m L = 1 \,\,\,\, (\times 4) & , \\[2mm]
ii) & \varphi = \pm \tilde{\varphi} \neq 0  &  \Rightarrow &  m L = 1 \,\,\,\, (\times 3) \,\,\,\, , \,\,\,\, 3 - \dfrac{2}{1+\tilde{\varphi}^2} & , \\[4mm]
iii) & \varphi = 0 &  \Rightarrow &  m L = 1 \,\,\,\, (\times 2) \,\,\,\, , \,\,\,\, \sqrt{1+\tilde{\varphi}^2}  \,\,\,\, (\times 2) & , \\[3mm]
iv) & \tilde{\varphi} = 0 &  \Rightarrow &  m L = 1 \,\,\,\, (\times 2) \,\,\,\, , \,\,\,\, \sqrt{1+\varphi^2}  \,\,\,\, (\times 2) & .
\end{array}    
\end{equation}
Note that the case $\,i)\,$ sits at the intersection of the one-dimensional slicings $\,ii)\,$, $\,iii)\,$ and $\,iv)\,$. A diagram of the $\,(\varphi \,;\tilde{\varphi})$-family of AdS$_{4}$ solutions in (\ref{vacuum_N=4_N=2}) is shown in Figure~\ref{fig:diagram}. In the figure, and in the rest of the work, we have denoted by $\,\mathcal{N} \, \& \, \textrm{G}_{0}\,$ the number $\,\mathcal{N}\,$ of four-dimensional supersymmetries and the residual symmetry group $\,\textrm{G}_{0}\,$ of a given AdS$_{4}$ solution.

\subsubsection{$\mathcal{N}=3\,$ line of supersymmetry enhancement}

\begin{figure}[t!]
\begin{center}
\includegraphics[width=0.9\textwidth]{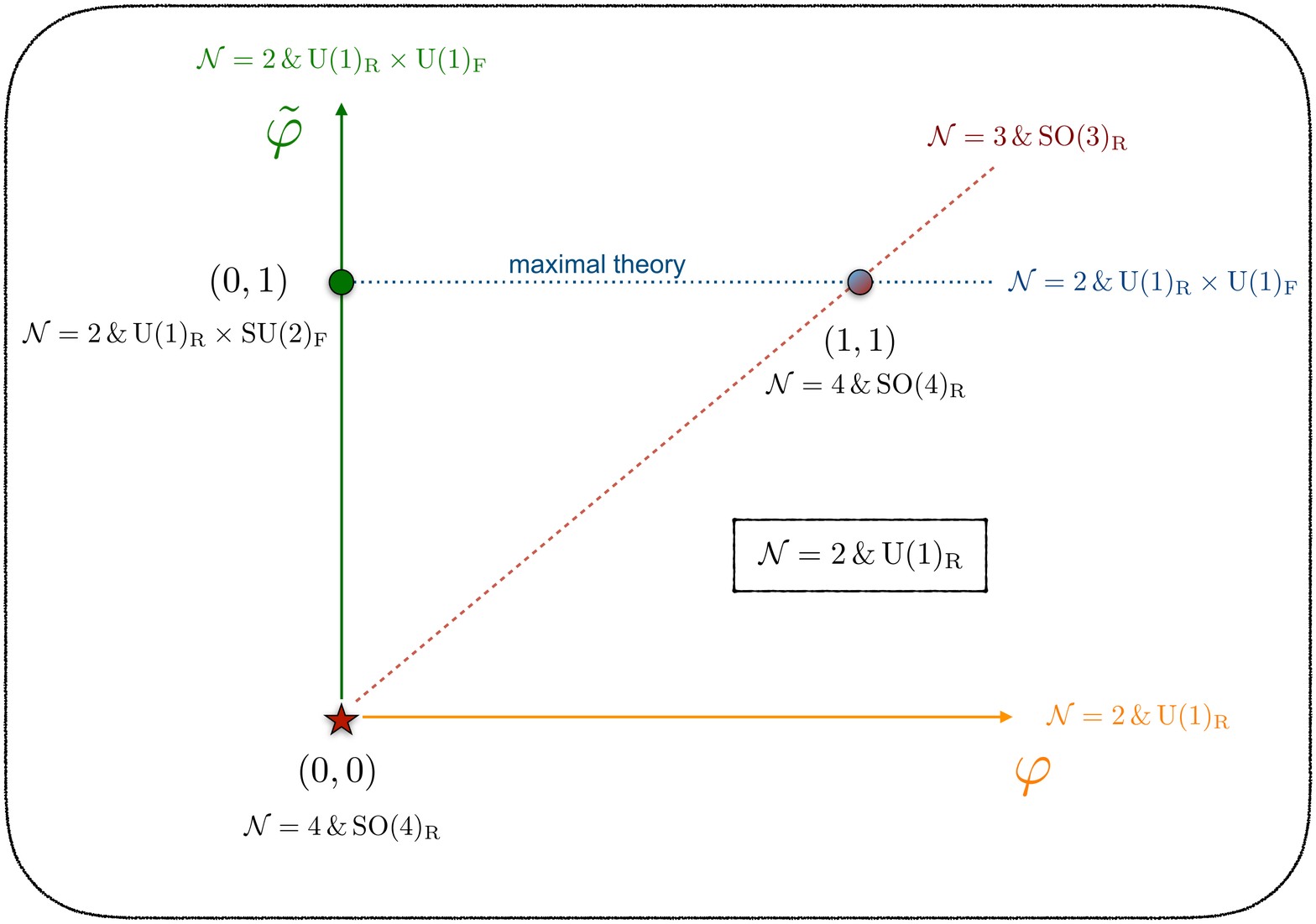}
\put(-58,177){\cite{Bobev:2021yya}} 
\put(-146,147){\cite{Inverso:2016eet}} 
\put(-350,161){\cite{Guarino:2020gfe}}
\put(-303,14){\cite{Dibitetto:2011gm}} 
\caption{Web of $\,\mathcal{N}=2\,$ CFT$_{3}$'s dual to the $\,(\varphi \,;\tilde{\varphi})$-family of AdS$_{4}$ solutions in (\ref{vacuum_N=4_N=2}). CFT$_{3}$'s at $\,{\tilde{\varphi} = 1}\,$ (dotted blue line) are dual to AdS$_{4}$ solutions of the $\,[\textrm{SO}(1,1) \times \textrm{SO}(6)] \ltimes \mathbb{R}^{12}\,$ maximal supergravity and have been studied and characterised in \cite{Bobev:2021yya}. The green and blue/red circles have a type IIB dual: the S-fold backgrounds in \cite{Guarino:2020gfe} and \cite{Inverso:2016eet}, respectively. The $\mathcal{N}=4\,$ CFT$_{3}$ sitting at $\,\varphi = \tilde{\varphi} = 0\,$ (red star) is dual to an exotic AdS$_{4}$ solution originally presented in \cite{Dibitetto:2011gm} and classified as non-geometric therein. A generic CFT$_{3}$ in the diagram features $\,\mathcal{N}=2 \,\& \, \textrm{U}(1)_{\textrm{R}}\,$ symmetry.}
\label{fig:diagram}
\end{center}
\end{figure}

The two involutions $\,\varphi = \pm \tilde{\varphi}\,$ respectively yield $\,\Delta_\mp=\frac{3}{2}\,$ so that the corresponding long multiplet in (\ref{long_multiplets_general_Z2xZ2}) hits the unitary bound. We will set $\,\varphi = \tilde{\varphi}\,$ (red dashed line in Figure~\ref{fig:diagram}) without loss of generality by virtue of the reflection symmetry $\,\tilde{\varphi} \rightarrow -\tilde{\varphi}\,$ of (\ref{vacuum_N=4_N=2}) and (\ref{Deltas_general_Z2xZ2}).

The conformal dimensions in (\ref{Deltas_general_Z2xZ2}) reduce in this case to\footnote{One has that $\,\tilde{\Delta}_{-}=2\,$ and $\,\tilde{\Delta}_{+}=3-\tfrac{2}{1+\tilde{\varphi}^2}\,$ for $\,|\tilde{\varphi}| \ge 1\,$ whereas $\,\tilde{\Delta}_{+}=2\,$ and $\,\tilde{\Delta}_{-}=3-\tfrac{2}{1+\tilde{\varphi}^2}\,$ for $\,|\tilde{\varphi}| \le 1\,$.}
\begin{equation}
\label{Deltas_N=3_Z2xZ2}
\Delta_{1} = 4-\tfrac{2}{1+\tilde{\varphi}^2}
\hspace{4mm} , \hspace{4mm}
\Delta_{-}=\tfrac{3}{2}
\hspace{4mm} , \hspace{4mm}
\Delta_{+} = \tfrac{7}{2}-\tfrac{2}{1+\tilde{\varphi}^2}
\hspace{4mm} , \hspace{4mm}
\tilde{\Delta}_{-} = 2
\hspace{4mm} , \hspace{4mm}
\tilde{\Delta}_{+} = 3-\tfrac{2}{1+\tilde{\varphi}^2} \ .
\end{equation}
As a result, the long multiplet $\,L\bar{L}[\tfrac{1}{2}]^{0}_{\Delta_{-}}\,$ hits the unitarity bound and splits into one short and two semi-short multiplets. The multiplets in (\ref{long_multiplets_general_Z2xZ2}) then reduce to
\begin{equation}
\label{long_multiplets_N=3_Z2xZ2_osp2|4}
\begin{array}{rcl}
     L\bar{L}[0]^{0}_{\Delta_1}  & \to & L\bar{L}[0]^{0}_{4-\frac{2}{1+\tilde{\varphi}^2}}  \\[6mm]
     L\bar{L}[\tfrac{1}{2}]^{0}_{\Delta_{-}} \,\,\oplus\,\,  L\bar{L}[\tfrac{1}{2}]^{0}_{\Delta_{+}} & \to & \left[  A_1\bar{A_1}[\tfrac{1}{2}]_{\frac{3}{2}}^{0} \,\,\oplus\,\, A_2\bar{L}[0]^{-1}_2 \,\,\oplus\,\, L\bar{A_2}[0]^{1}_2 \right] \,\,\oplus\,\, L\bar{L}[\tfrac{1}{2}]^{0}_{\frac{7}{2}-\frac{2}{1+\tilde{\varphi}^2}}
     \\[6mm]
    L\bar{L}[0]_{\tilde{\Delta}_{-}}^{0} \,\, \oplus\,\, L\bar{L}[0]_{\tilde{\Delta}_{+}}^{0} &\to& L\bar{L}[0]_{2}^{0} \,\,\oplus\,\,  L\bar{L}[0]_{3-\frac{2}{1+\tilde{\varphi}^2}}^{0}
\end{array}
\end{equation}
The enhancement to $\,\mathcal{N}=3\,$ supersymmetry originates from the short multiplet  $\,A_1\bar{A_1}[\tfrac{1}{2}]_{\frac{3}{2}}^{0}\,$ in (\ref{long_multiplets_N=3_Z2xZ2_osp2|4}) which contains a massless gravitino.

Alternatively, the mass spectrum in (\ref{short&semi-short_multiplets_general_Z2xZ2}) and (\ref{long_multiplets_N=3_Z2xZ2_osp2|4}) can be arranged into unitary superconformal multiplets of the $\,\mathfrak{osp}(3|4)\,$ algebra. Following closely the notation\footnote{Our convention for the  Lorentz and R-symmetry Dynkin labels differs from the one in \cite{Cordova:2016emh}: $\,j=\frac{1}{2} j_{\tiny{\cite{Cordova:2016emh}}}\,$ and $\,R=\frac{1}{2} R_{\tiny{\cite{Cordova:2016emh}}}\,$.} of \cite{Cordova:2016emh}, we find a set of multiplets
\begin{equation}
\label{multiplets_N=3_Z2xZ2}
L[0]_{3-\frac{2}{1 + \tilde{\varphi}^2}}^{0} 
\hspace{5mm} , \hspace{5mm}
B_{1}[0]_{2}^{2} 
\hspace{5mm} , \hspace{5mm}
A_{1}[\tfrac{1}{2}]_{\frac{3}{2}}^{0}  \ ,
\end{equation}
with $\,A_{1}[\tfrac{1}{2}]_{\frac{3}{2}}^{0} \,$ corresponding to the stress-energy multiplet of the dual $\,\mathcal{N}=3\,$ CFT$_{3}$. The unprotected long multiplet in (\ref{multiplets_N=3_Z2xZ2}) is simply a rearrangement of the $\,\tilde{\varphi}$-dependent long multiplets in (\ref{long_multiplets_N=3_Z2xZ2_osp2|4}).

\subsubsection{$\mathcal{N}=4\,$ points of supersymmetry enhancement}

There are two isolated points in the space of conformal dimensions (\ref{Deltas_N=3_Z2xZ2}) at which $\,\Delta_{1}\,$ and $\,\tilde{\Delta}_{\pm}\,$ corresponding to a $\,[j]=[0]\,$ HWS are integer valued whereas $\,\Delta_{\pm}\,$ corresponding to a $\,[j]=[\frac{1}{2}]\,$ HWS are half-integer valued.\footnote{There is also the limit $\,\tilde{\varphi} \rightarrow \pm \infty\,$ for which the gauging changes drastically pointing at a decompactification regime (see discussion below (\ref{algebra_ISO3_1_family_I})-(\ref{algebra_ISO3_2_family_I})). Studying this limit goes beyond the scope of this work.} Let us look at these two special points in more detail.

\paragraph{$\bullet$ Point $\,\varphi=\tilde{\varphi}=1\,$:} 

At this point (red/blue circle in Figure~\ref{fig:diagram}), the conformal dimensions in (\ref{Deltas_N=3_Z2xZ2}) simplify to
\begin{equation}
\label{Deltas_N=4_S-fold_Family_I}
\Delta_{1} = 3
\hspace{5mm} , \hspace{5mm}
\Delta_{-}=\tfrac{3}{2} 
\hspace{5mm} , \hspace{5mm}
\Delta_{+}=\tfrac{5}{2} 
\hspace{5mm} , \hspace{5mm}
\tilde{\Delta}_{\pm}=2 \ .
\end{equation}
Therefore, there are no further long multiplets in (\ref{long_multiplets_N=3_Z2xZ2_osp2|4}) hitting the unitarity bound. Instead, they simply reduce to
\begin{equation}
\label{splitting_long_N=4_S-fold_Z2xZ2}
\begin{array}{rcl}
     L\bar{L}[0]^{0}_{\Delta_1}  & \to & L\bar{L}[0]^{0}_{3}  \\[4mm]
     L\bar{L}[\tfrac{1}{2}]^{0}_{\Delta_{-}} \,\,\oplus\,\,  L\bar{L}[\tfrac{1}{2}]^{0}_{\Delta_{+}} & \to & \left[  A_1\bar{A_1}[\tfrac{1}{2}]_{\frac{3}{2}}^{0} \,\,\oplus\,\, A_2\bar{L}[0]^{-1}_2 \,\,\oplus\,\, L\bar{A_2}[0]^{1}_2 \right] \,\,\oplus\,\, L\bar{L}[\tfrac{1}{2}]^{0}_{\frac{5}{2}}
     \\[6mm]
    L\bar{L}[0]_{\tilde{\Delta}_{-}}^{0} \,\, \oplus\,\, L\bar{L}[0]_{\tilde{\Delta}_{+}}^{0} &\to& L\bar{L}[0]_{2}^{0}\,\,\,\, (\times 2)
\end{array}
\end{equation}
The multiplets in (\ref{short&semi-short_multiplets_general_Z2xZ2}) and (\ref{splitting_long_N=4_S-fold_Z2xZ2}) precisely describe the $\mathbb{Z}_{2}$-even sector of the $\,{\mathcal{N}=4 \, \& \, \textrm{SO}(4)}_{\textrm{R}}\,$ S-fold of \cite{Inverso:2016eet}. In other words, the field content of half-maximal supergravity only captures an $\,\mathcal{N}=3\,$ subsector of the $\,\mathcal{N}=4\,$ S-fold CFT$_{3}$ of \cite{Inverso:2016eet}.

In $\,\mathfrak{osp}(3|4)\,$ language, the unprotected long multiplet in (\ref{multiplets_N=3_Z2xZ2}) does not split and the spectrum (\ref{multiplets_N=3_Z2xZ2}) simply reduces to
\begin{equation}
\label{N=4_S-fold_Z2_modding}
L[0]_{2}^{0} 
\hspace{5mm} , \hspace{5mm}
B_{1}[0]_{2}^{2} 
\hspace{5mm} , \hspace{5mm}
A_{1}[\tfrac{1}{2}]_{\frac{3}{2}}^{0}   \ .
\end{equation}

\paragraph{$\bullet$ Point $\,\varphi=\tilde{\varphi}=0\,$:} At this special point (red star in Figure~\ref{fig:diagram}) the conformal dimensions in (\ref{Deltas_N=3_Z2xZ2}) reduce to
\begin{equation}
\label{Deltas_N=4__Z2xZ2}
\Delta_{1} = 2
\hspace{5mm} , \hspace{5mm}
\Delta_{\pm}=\frac{3}{2} 
\hspace{5mm} , \hspace{5mm}
\tilde{\Delta}_{-}=2
\hspace{5mm} , \hspace{5mm}
\tilde{\Delta}_{+}=1 \ .
\end{equation}
This implies that the three long multiplets $\,L\bar{L}[\tfrac{1}{2}]^{0}_{\Delta_{\pm}}\,$ and $\,L\bar{L}[0]^{0}_{\tilde{\Delta}_{+}}\,$ hit the unitarity bound and split, each of them producing one short and two semi-short multiplets. More concretely, the multiplets in (\ref{long_multiplets_N=3_Z2xZ2_osp2|4}) decompose as
\begin{equation}
\label{long_multiplets_N=4_Z2xZ2_osp2|4}
\begin{array}{rcl}
     L\bar{L}[0]^{0}_{\Delta_1}  &\to &  L\bar{L}[0]^{0}_{2} \\[4mm]
     L\bar{L}[\tfrac{1}{2}]^{0}_{\Delta_{\pm}} & \to &\left[  A_1\bar{A_1}[\tfrac{1}{2}]_{\frac{3}{2}}^{0} \,\,\oplus\,\, A_2\bar{L}[0]^{-1}_2 \,\,\oplus\,\, L\bar{A_2}[0]^{1}_2 \right]  \,\,\,\, (\times 2)  \\[6mm]
     L\bar{L}[0]_{\tilde{\Delta}_{-}}^{0} \,\,\oplus\,\,  L\bar{L}[0]_{\tilde{\Delta}_{+}}^{0} & \to & L\bar{L}[0]_{2}^{0} \,\,\oplus\,\,   \Big[ A_2\bar{A_2}[0]^{0}_1 \,\,\oplus\,\, B_1\bar{L}[0]^{-2}_2 \,\,\oplus\,\,   L\bar{B_1}[0]^{2}_2 \Big]
\end{array}
\end{equation}
There is this time an enhancement to $\,\mathcal{N}=4\,$ supersymmetry originating from the two short multiplets  $\,A_1\bar{A_1}[\tfrac{1}{2}]_{\frac{3}{2}}^{0}\,$ in (\ref{long_multiplets_N=4_Z2xZ2_osp2|4}) each one containing a massless gravitino. In addition, there is the shortening associated with $\,\tilde{\Delta}_{+}=1\,$ which provides an additional massless vector multiplet.

In $\,\mathfrak{osp}(3|4)\,$ language, the unprotected long multiplet in (\ref{multiplets_N=3_Z2xZ2}) hits the unitarity bound and splits into two short multiplets as $\,L[0]^0_1 \,\rightarrow\, A_2[0]^0_1 \,\oplus B_1[0]^2_2\,$. The spectrum in (\ref{multiplets_N=3_Z2xZ2}) then reduces to 
\begin{equation}
A_2[0]^0_1
\hspace{5mm} , \hspace{5mm}
B_1[0]^2_2 \,\,\,\, (\times 2)
\hspace{5mm} , \hspace{5mm}
A_{1}[\tfrac{1}{2}]_{\frac{3}{2}}^{0} \ ,
\end{equation}
as a consequence of the supersymmetry enhancement to $\,\mathcal{N}=4\,$ in the exotic CFT$_{3}$.

\subsubsection{$\textrm{U}(1)_{\textrm{F}}\,$ flavour symmetry enhancement}

The identification $\,\varphi =0\,$ (vertical axis in Figure~\ref{fig:diagram}) gives rise to a $\,\textrm{U}(1)_{\textrm{F}}\,$ flavour symmetry enhancement in the corresponding CFT$_{3}$'s. At this value it occurs that $\,\tilde{\Delta}_{-}=1\,$, the long multiplet $\,L\bar{L}[0]_{\tilde{\Delta}_{-}}^0\,$ in (\ref{long_multiplets_general_Z2xZ2}) hits the unitarity bound and splits again into one short and two semi-short multiplets. The multiplets in (\ref{long_multiplets_general_Z2xZ2}) reduce to
\begin{equation}
\label{N=2_long_splitting_varphi=0_varphitilde=0_Z2xZ2}
\begin{array}{rcl}
     L\bar{L}[0]^{0}_{\Delta_1}  & \to & L\bar{L}[0]^{0}_{\tfrac{1}{2}+\tfrac{1}{2}\sqrt{25-\frac{16}{1 + \tilde{\varphi}^2}}} \\[6mm]
     L\bar{L}[\tfrac{1}{2}]^{0}_{\Delta_{\pm}} & \to & L\bar{L}[\tfrac{1}{2}]^0_{
     1+\frac{1}{2}\sqrt{5+4\tilde{\varphi}^2-4 \sqrt{\tilde{\varphi}^2+1}}}\,\,\,\,   (\times 2)  \\[6mm]
     L\bar{L}[0]_{\tilde{\Delta}_{-}}^{0} \,\,\oplus\,\,  L\bar{L}[0]_{\tilde{\Delta}_{+}}^{0} & \to & \Big[ A_2\bar{A}_2[0]^0_1 \,\, \oplus  \,\, L\bar{B}_1[0]_2^2 \,\, \oplus  \,\, B_1\bar{L}[0]^{-2}_2 \Big] \,\, \oplus \,\, L\bar{L}[0]^{0}_{\tfrac{1}{2}+\tfrac{1}{2}\sqrt{9 + \frac{16 \, \tilde{\varphi}^4}{1 + \tilde{\varphi}^2}}} 
\end{array}
\end{equation}
where $\,A_2\bar{A}_2[0]^0_1\,$ is a massless vector multiplet reflecting the $\,\textrm{U}(1)_{\textrm{F}}\,$ flavour symmetry enhancement in the dual $\,\mathcal{N}=2\,$ CFT$_{3}$'s. 

Note that the two degenerated long multiplets   $\,L\bar{L}[\tfrac{1}{2}]^{0}_{\Delta_{\pm}}\,$ in (\ref{N=2_long_splitting_varphi=0_varphitilde=0_Z2xZ2}) hit the unitarity bound at the special point $\,\tilde{\varphi}=0\,$. At this point, each of them splits as
\begin{equation}
\begin{array}{llllllll}
\label{splitting_phitilde=0_family_I}
L\bar{L}[\tfrac{1}{2}]_{\frac{3}{2}}^0 & \rightarrow & A_1\bar{A}_1[1]^0_{\frac{3}{2}} & \oplus & A_2 \bar{L}[0]^{-1}_{2} & \oplus & L\bar{A}_2[0]^{1}_2 
\end{array}
\end{equation}
recovering the exotic $\,\mathcal{N}=4\,$ CFT$_{3}$.

\subsection{General axion deformations of the exotic $\mathcal{N}=4\,$ CFT$_{3}$}
\label{sec:chi_def_exotic_N=4}

Setting $\,\varphi=\tilde{\varphi}=0\,$ in (\ref{vacuum_N=4_N=2}) reproduces the exotic $\,\mathcal{N}=4\,$ AdS$_{4}$ solution with $\,\textrm{SO}(4)_{\textrm{R}}\,$ symmetry originally reported in \cite{Dibitetto:2011gm} (red star in Figure~\ref{fig:diagram}). Let us recall that it is located at the (rescaled by $c$) origin of the scalar geometry, namely, 
\begin{equation}
\label{vacuum_N=4_chi=0}
z_{1,2,3} = i \, c \, \hspace{10mm} , \hspace{10mm}
z_{4,5,6,7}= i  \ ,
\end{equation}
and preserves the compact part of the gauging, namely, 
\begin{equation}
\textrm{SO}(4)_{\textrm{R}} \,\sim\, \textrm{SO}(3)_{1} \times \textrm{SO}(3)_{2} \,\subset\, \textrm{ISO}(3)_{1} \times \textrm{ISO}(3)_{2}   \ .
\end{equation}
Three new scalar moduli can be turned on at this AdS$_{4}$ solution which are parameterised by three axions $\,\chi_{1,2,3} \in \mathbb{R}\,$ of the axion-like type investigated for the S-fold solutions in \cite{Giambrone:2021zvp,Guarino:2021kyp,Guarino:2021hrc,Guarino:2022tlw}. Activating $\,\chi_{1,2,3}\,$ changes the location of the AdS$_{4}$ solution (\ref{vacuum_N=4_chi=0}) to
\begin{equation}
\label{vacuum_N=4_chi}
z_{1,2,3} = c \,\, (-\chi_{1,2,3} + i) \hspace{10mm} , \hspace{10mm}
z_{4,5,6,7}= i  \ ,
\end{equation}
keeping the vacuum energy at the value $\,V_{0}= - 3 \, g^2 c^{-1}$. Therefore, the moduli fields $\,\chi_{i}\,$ ($i=1,2,3$) are naturally identified with marginal deformations of the $\,\mathcal{N}=4\,$ exotic CFT$_{3}$.

The explicit computation of the normalised gravitino masses at the AdS$_{4}$ solution (\ref{vacuum_N=4_chi}) yields
\begin{equation}
m L \,\,=\,\, 
\sqrt{1 + \omega_{1}^2}
\hspace{4mm} , \hspace{4mm}
\sqrt{1 + \omega_{2}^2}
\hspace{4mm} , \hspace{4mm}
\sqrt{1 + \omega_{3}^2}
\hspace{4mm} , \hspace{4mm}
\sqrt{1 + \omega_{4}^2} \ ,
\end{equation}
with 
\begin{equation}
\label{chi&omega_def}
\chi_{i} = \omega_{j} + \omega_{k} \hspace{5mm} (i \neq j \neq k) 
\hspace{10mm} \textrm{ and } \hspace{10mm}\omega_{1}+\omega_{2}+\omega_{3} +\omega_4 = 0 \ .
\end{equation}
It proves very convenient to introduce a set $\,\omega_{A}=\{\omega_{1},\omega_{2},\omega_{3},\omega_{4}\}\,$ of deformation parameters subject to the constraint $\,\sum_{A=1}^{4}\omega_{A}=0$. The twelve normalised vector masses can then be very symmetrically written as
\begin{equation}
\label{chi_exotic_vector_masses}
m^2 L^2 = 1 + \omega_A^2 + \omega_B^2 \pm \sqrt{ 1 +(\omega_A+\omega_B)^2+ 4\, \omega_A^2\, \omega_B^2}
\hspace{8mm} \textrm{ with } \hspace{8mm}
A < B \ ,
\end{equation}
whereas the normalised scalar masses are given by
\begin{equation}
\label{scalar_masses_omega}
\begin{array}{rcl}
m^2 L^2  & = & -1 + \omega_A^2 + \omega_B^2 \pm \sqrt{1 + (\omega_A + \omega_B)^2 + 4 \, \omega_A^2 \, \omega_B^2}\hspace{8mm} \textrm{ with } \hspace{8mm} A < B \ , \\[3mm]
& & 4 \,\,\,,\,\,\,-2\,\,\, (\times 2)\,\,\,,\,\,\,0\,\,\, (\times 15)\,\,\,,\,\,\,\lambda_{1,\ldots,8} \,\,\, (\times 1) \ .
\end{array}
\end{equation}
We cannot provide a closed form for the eight normalised scalar masses $\,\lambda_{1,\ldots,8}\,$. They correspond to the eigenvalues of the $\,\chi_{1,2,3}$-dependent matrix
\begin{equation}
\small{
   \left(
\begin{array}{cccccccc}
 \hat{\chi}^2-2 & 0 & 2 \chi _2 \chi _3 & 2 \chi _1 \chi _3 & 2 \chi _1 \chi _2 & 0 & 0 & 0 \\
 0 & \hat{\chi} ^2-2 & 2 \chi _1 & 2 \chi _2 & 2 \chi _3 & 2 \chi _2^2-\hat{\chi} ^2 & 2 \chi _3^2-\hat{\chi}^2 & 2 \chi _1^2-\hat{\chi} ^2 \\
 2 \chi _2 \chi _3 & 2 \chi _1 & \hat{\chi}^2 & 2 \chi _1 \chi _2 & 2 \chi _1 \chi _3 & -2 \chi _1 & -2 \chi _1 & 0 \\
 2 \chi _1 \chi _3 & 2 \chi _2 & 2 \chi _1 \chi _2 & \hat{\chi}^2 & 2 \chi _2 \chi _3 & 0 & -2 \chi _2 & -2 \chi _2 \\
 2 \chi _1 \chi _2 & 2 \chi _3 & 2 \chi _1 \chi _3 & 2 \chi _2 \chi _3 & \hat{\chi}^2 & -2 \chi _3 & 0 & -2 \chi _3 \\
 0 & 2 \chi _2^2-\hat{\chi} ^2 & -2 \chi _1 & 0 & -2 \chi _3 & \hat{\chi}^2 & 2-\hat{\chi} ^2+2 \chi _1^2 & 2-\hat{\chi}^2+2 \chi _3^2 \\
 0 & 2 \chi _3^2-\hat{\chi}^2 & -2 \chi _1 & -2 \chi _2 & 0 & 2-\hat{\chi}^2+2 \chi _1^2 & \hat{\chi}^2 & 2 -\hat{\chi}^2+2 \chi _2^2 \\
 0 & 2 \chi _1^2-\hat{\chi}^2 & 0 & -2 \chi _2 & -2 \chi _3 & 2 -\hat{\chi}^2+2 \chi _3^2 & 2-\hat{\chi}^2+2 \chi _2^2 & \hat{\chi}^2 \\
\end{array}
\right) ,}
\end{equation}
with $\,\hat{\chi}^2 \equiv \chi_{1}^2+\chi_{2}^2+\chi_{3}^2\,$. A numerical scan of the mass spectrum in \eqref{scalar_masses_omega} as a function of $\,\omega_{A}\,$ shows that there exist regions in parameter space for which non-supersymmetric solutions become perturbatively unstable, \textit{e.g.} whenever $\,\omega_1 = \omega_2 = - \omega_3 = -\omega_4 > 0.3052\,$.

It follows from (\ref{chi_exotic_vector_masses}) that the number of massless vectors is given by the number $\,n_{p}\,$ of pairs $\,\omega_A = \omega_B\,$ with $\,A<B\,$. This yields the following classification of CFT$_{3}$ duals in terms of the four parameters $\,\omega_{A}\,$:
\begin{itemize}
\item[$\circ$] All four $\,\omega$'s\, are zero: $\mathcal{N}=4$ with $\,\textrm{SO}(4)_{\textrm{R}}\,$ symmetry
\item[$\circ$] Three $\,\omega$'s\, are zero: same case as before by virtue of $\,\sum \omega_{A}=0$
\item[$\circ$] Exactly two $\,\omega$'s\, are zero: $\,\mathcal{N}=2\,$ with $\,\textrm{SO}(2)_\textrm{R}\,$ symmetry
\item[$\circ$] Exactly one $\,\omega\,$ is zero: $\,\mathcal{N}=1\,$ with
    \begin{itemize}
        \item $\textrm{SO}(2)_\textrm{F}\,$ flavour symmetry if two of the remaining $\,\omega$'s\, are equal
        \item no flavour symmetry otherwise
        \end{itemize}
\item[$\circ$] All $\,\omega$'s\, are non-zero: $\mathcal{N}=0\,$ with
    \begin{itemize}
        \item $\textrm{SO}(2)_{\textrm{F}}\times \textrm{SO}(2)_{\textrm{F}}\,$ if there are two pairs of $\omega$'s\, of equal value, one pair with the opposite sign of the other
        \item $\textrm{SO}(3)_{\textrm{F}}\,$ if three $\,\omega$'s\, are identified
        \item $\textrm{SO}(2)_{\textrm{F}}$ if only two $\,\omega$'s\, are identified
        \item no flavour symmetry otherwise
    \end{itemize}
\end{itemize}
In summary, the number of supersymmetries (both of the AdS$_{4}$ solution and the dual CFT$_{3}$) matches the number of parameters $\,\omega_A = 0$, and the AdS$_{4}$ solution features an orthogonal symmetry group of dimension $\,n_{p}$. Amongst the $\,n_{p}\,$ pairs, each pair $\,\omega_{A}=\omega_{B}=0\,$ adds one generator to the orthogonal R-symmetry group of the dual CFT$_{3}$ whereas each pair $\,\omega_{A}=\omega_{B}\neq 0\,$ adds one generator to an orthogonal flavour symmetry group in the CFT$_{3}$. Finally, observe that there is a case compatible with a single axion, let us denote it $\,\chi$, that preserves $\,\mathcal{N}=2\,$ and $\,\textrm{SO}(2)_{\textrm{R}}\,$ symmetry within half-maximal supergravity. The consequences of turning on this axion $\,\chi\,$ will be investigated in detail in Section~\ref{sec:U(1)_sector}.

\subsubsection{Relation to axion-like deformations of S-folds?}
\label{sec:exotic_chi's}

Axion-like deformations (generically also denoted by $\,\chi$'s) of S-fold solutions of type IIB supergravity have been geometrically characterised in \cite{Guarino:2021kyp,Guarino:2021hrc} (see \cite{Guarino:2022tlw} for a light review). In these solutions, they cause a breaking of symmetries by $\textit{global}$ effects in the ten dimensional geometry \cite{Giambrone:2021zvp,Guarino:2021kyp}. More concretely, axion-like deformations break the symmetry group of an S-fold down to its Cartan subgroup. The number of such deformations was shown to match the dimension of the Cartan subalgebra of the symmetry group of the S-fold solution before turning on any deformation.

The way the $\,\chi$'s enter the AdS$_{4}$ solution in (\ref{vacuum_N=4_chi}) is identical to the known examples of axion-like deformations of S-folds (\textit{cf.} eq.$(2.18)$ in \cite{Guarino:2020gfe}). However one must be cautious about giving the same interpretation to all the $\,\chi$'s in (\ref{vacuum_N=4_chi}). The reason is twofold:
\begin{itemize}

\item[$i)\,$] We find three different axion deformations $\,\chi_{1,2,3}\,$ for the exotic $\,\mathcal{N}=4\,$ AdS$_{4}$ solution with $\,\textrm{SO}(4)_{\textrm{R}}\,$ symmetry and not two, as one would naively expect from the number of Cartan generators of $\,\textrm{SO}(4)_{\textrm{R}}\,$.

\item[$ii)\,$] Generic values of $\,\chi_{1,2,3}\,$ totally break the $\,\textrm{SO}(4)_{\textrm{R}}\,$ symmetry of the undeformed solution: the Cartan subgroup of $\,\textrm{SO}(4)_{\textrm{R}}\,$ is generically not preserved. As a result, the pattern of symmetry breaking is very different from the one induced by the axion-like deformations in the S-fold solutions \cite{Guarino:2021kyp,Guarino:2021hrc}. 

\end{itemize}

As a result, while the $\,\mathcal{N}=0\,$ case with $\,\textrm{SO}(2)_{\textrm{F}}\times \textrm{SO}(2)_{\textrm{F}}\,$ symmetry preserves the Cartan subgroup of $\,\textrm{SO}(4)_{\textrm{R}}\,$ and, therefore, stands a chance of having a geometrical interpretation alike the axion-like deformations of S-folds, the other cases appear (at first glance) to be different as the Cartan subgroup of $\,\textrm{SO}(4)_{\textrm{R}}\,$ is not preserved. Of special interest will be the case of the single axion $\,\chi\,$ mentioned above, which we move to discuss in the next section. This axion preserves $\,\mathcal{N}=2\,$ supersymmetry, an $\,\textrm{SO}(2)_{\textrm{R}} \subset \textrm{SO}(4)_{\textrm{R}}\,$ symmetry within half-maximal supergravity, and combines with the modulus $\,\varphi\,$ to generalise the conformal manifold of $\,\mathcal{N}=2\,$ CFT$_{3}$'s in (\ref{CM_metric_intro}) to arbitrary values of the parameter $\,\tilde{\varphi}$.

\section{$\textrm{U}(1)_{\textrm{R}}$-invariant sector}
\label{sec:U(1)_sector}

In this section we construct a particular $\,\textrm{U}(1)_\textrm{R}\,$ invariant sector of half-maximal supergravity that suffices to capture the modulus $\,\chi\,$ dual to the second marginal deformation spanning with $\,\varphi\,$ the $\,\mathcal{N}=2\,$ conformal manifold in (\ref{CM_metric_intro}).

\subsection{The $\,\mathcal{N}=2\,$ three-vector and two-hyper model}
\label{sec:N=2_model}

To our knowledge, there is no explicit construction of the $\,\textrm{U}(1)_\textrm{R}\,$ invariant sector of half-maximal supergravity of relevance for this work. So we will present it in some detail. In order to construct it, let us first introduce the set of $\,\textrm{SO}(6,6)\,$ generators
\begin{equation}
\left[t_{MN} \right]_P{}^Q = 2 \,  \eta_{P[M} \, \delta^Q_{N]} \ ,
\end{equation}
where $\,M=1\ldots,6,\bar{1},\ldots\bar{6}\,$ is a fundamental $\,\textrm{SO}(6,6)\,$ index in the light-cone basis. We choose the specific $\,\textrm{U}(1)_\textrm{R}\,$ generator to be
\begin{equation}
\label{t_U(1)}
    t_{\textrm{U}(1)_\textrm{R}} = \left( t_{5\bar{1}} - t_{1\bar{5}} \right) - \left( t_{6\bar{2}} - t_{2\bar{6}}  \right) \ ,
\end{equation}
which is embedded in the duality group of half-maximal supergravity as
\begin{equation}
\label{U(1)_R_embedding}
\textrm{SL}(2) \times \textrm{SO}(6,6) \supset \textrm{SL}(2) \times \textrm{SO}(2,2) \times \textrm{SO}(4,4) \supset \textrm{SL}(2) \times \textrm{SO}(2,2) \times \textrm{SU}(2,2) \times \textrm{U}(1)_{\textrm{R}} \ .
\end{equation}
From the commutant of $\,\textrm{U}(1)_{\textrm{R}}\,$ in the embedding chain (\ref{U(1)_R_embedding}), the scalar manifold invariant under $\,\textrm{U}(1)_\textrm{R}\,$ is identified with
\begin{equation}
\label{scalar_geometry_N2}
\mathcal{M}^{^{\textrm{U}(1)_{\textrm{R}}}}_{\textrm{scal}} = \underbrace{\frac{\textrm{SL}(2)}{\textrm{SO}(2)} \times 
\frac{\textrm{SO}(2,2)}{\textrm{SO}(2)\times \textrm{SO}(2)}}_{\left[ \dfrac{ \textrm{SL}(2)}{\textrm{SO}(2)}  \right]^3}
\times \frac{\textrm{SU}(2,2)}{\textrm{S}\left(\textrm{U}(2)\times \textrm{U}(2) \right)} \ .
\end{equation}

This $\,\textrm{U}(1)_\textrm{R}$-invariant sector of half-maximal supergravity can be described as an $\,\mathcal{N}=2\,$ gauged supergravity coupled to three vector multiplets and two hyper-multiplets. Within this $\,\mathcal{N}=2\,$ sector, the gauge group is
\begin{equation}
\label{G_N=2}
\textrm{G}_{\mathcal{N}=2} = \textrm{U}(1)_\textrm{R} \times \textrm{U}(1)_\gamma \times \mathbb{R}_a \times \mathbb{R}_\epsilon \subset \textrm{ISO}(3) \times  \textrm{ISO}(3) \ ,
\end{equation}
with all the fields being inert under the $\,\textrm{U}(1)_\textrm{R}\,$ factor. The three complex scalars in the vector multiplets span the $\,\left[\textrm{SL}(2)/\textrm{SO}(2)\right]^{3}\,$ factor of the scalar geometry (\ref{scalar_geometry_N2}). They are identified with $\,z_{7}\,$ and $\,(z_{2},z_{5})\,$ in the $\,\mathbb{Z}_2^2$-invariant sector of Section~\ref{sec:Z2xZ2_sector}. The scalar matrix $\,M_{\alpha\beta}\,$ spanned by $\,z_{7}\,$ was given in (\ref{M_SL2}). The part of the $\,M_{MN}\,$ matrix in (\ref{M_SO66}) spanned by $\,(z_{2},z_{5})\,$ was constructed in terms of the $\,2 \times 2\,$ blocks $\,G_{2}\,$ and $\,B_{2}\,$ in (\ref{G&B_blocks}). Alternatively, it can directly be constructed from the coset representative
\begin{equation}
\mathcal{V}_{\textrm{SO}(2,\,2)} =  e^{-\chi_2 \,  t_{4\bar{3}} + \chi_5 \,t_{43} } \,\, e^{ -\frac{1}{2} \,  \left[ \, 
\varphi_{2} \, (t_{4\bar{4}} - t_{3\bar{3}}) + \varphi_{5} (t_{4\bar{4}}+t_{3\bar{3}}) \, \right] } \ ,
\end{equation}
such that $\,z_{2,5}=-\chi_{2,5} + i e^{-\varphi_{2,5}}$. There is also the part of the scalar matrix $\,M_{MN}\,$ that depends on the scalars in the quaternionic K\"ahler (QK) space $\, \left[\textrm{SU}(2,2)/\textrm{S}(\textrm{U}(2) \times \textrm{U}(2)) \right] \sim \left[\textrm{SO}(2,4)/(\textrm{SO}(2) \times \textrm{SO}(4)) \right]$. Following the coset construction of \cite{Lu:1998xt}, we will first introduce the generators
\begin{equation}
\begin{array}{rclcrcl}
H_1 &=& t_{1\bar{1}} +t_{5\bar{5}} +t_{6\bar{6}} +t_{2\bar{2}}
& \hspace{5mm}, & \hspace{5mm}
H_2 &=& t_{1\bar{1}} +t_{5\bar{5}} -t_{6\bar{6}} -t_{2\bar{2}} \\[2mm]
{E_2}^3 &=& - t_{15}  
& \hspace{5mm}, & \hspace{5mm}
V^{23} &=& t_{26} \\[2mm]
U_1{}^3 &=&  -t_{12} + t_{56} + t_{16} -t_{25} 
& \hspace{5mm}, & \hspace{5mm}
U_1{}^2 &=& t_{2\bar{1}} + t_{2\bar{5}} + t_{6\bar{1}} - t_{6\bar{5}} \\[2mm]
U_2{}^3 &=& t_{12} - t_{56} + t_{16} -t_{25} 
& \hspace{5mm}, & \hspace{5mm}
U_2{}^2 &=& t_{2\bar{1}} - t_{2\bar{5}} - t_{6\bar{1}} - t_{6\bar{5}}
    \end{array}
\end{equation}
and construct the coset representative as
\begin{equation}
\mathcal{V}_{\textrm{SO}(2,4)} = e^{\frac{1}{2\sqrt{2}} \, U} \,\,\, e^{a \, V^{23}} \,\,\, e^{h \, E_{2}{}^{3}}
\,\,\, e^{-\frac{1}{4} \, \left[ \,  (\phi_{2}+\phi_{1}) H_{1} +  (\phi_{2} - \phi_{1}) H_{2} \, \right]} \ ,
\end{equation}
with
\begin{equation}
U = -(\tilde{\zeta}_{0}-\tilde{\zeta}_{1}) \, U_{1}{}^{2} -(\zeta^{0}+\zeta^{1}) \, U_{2}{}^{2} 
-(\zeta^{0}-\zeta^{1}) \, U_{1}{}^{3}
+(\tilde{\zeta}_{0}+\tilde{\zeta}_{1}) \, U_{2}{}^{3} \ .
\end{equation}
The scalar matrix $\,M_{MN} \in \textrm{SO}(6,6)\,$ is then obtained as
$\,M = \mathcal{V} \, \mathcal{V}^{t}\,$ using the factorised coset representative
\begin{equation}
\mathcal{V} = \mathcal{V}_{\textrm{SO}(2,2)} \,\, \mathcal{V}_{\textrm{SO}(2,4)} \ .
\end{equation}

In order to complete the characterisation of the $\,2+12\,$ scalars in the $\,\textrm{U}(1)_{\textrm{R}}$-invariant sector, we will now look at the metric on the scalar manifold (\ref{scalar_geometry_N2}) which can be extracted from the kinetic terms in (\ref{Lkin_general}). An explicit computation gives
\begin{equation}
\label{scalar_kinetic_SO(2,4)}
\begin{array}{rcl}
\mathcal{L}_{\textrm{kin}} &=& -\tfrac{1}{4} \left( d\varphi_{7}^2  + e^{2 \varphi_{7}}  \, d\chi_{7}^2 \right)  - \tfrac{1}{4} \left( d\varphi_{2}^2  + e^{2 \varphi_{2}}  \, d\chi_{2}^2 \right) - \tfrac{1}{4} \left( d\varphi_{5}^2  + e^{2 \varphi_{5}} \,  d\chi_{5}^2 \right) \\[3mm]
&-&  \tfrac{1}{4} \left( D\phi_{2}^2  + e^{2 \phi_{2}}  \, Dh^2 \right) \\[3mm]
&-& \frac{1}{4} \left[ d\phi_{1}^2 + e^{2\phi_{1} } \Big( Da + \tfrac12 \big( \zeta^0  D\tilde{\zeta}_0 + \zeta^1  D\tilde{\zeta}_1 - \tilde{\zeta}_0 D\zeta^0 - \tilde{\zeta}_1 D\zeta^1  \big) \Big)^2 \right] \\[4mm]
&-& \tfrac18 \left[ e^{\phi_{1} - \phi_{2}} \big( D\zeta^0 + D\zeta^1 \big)^2 
+ e^{\phi_{1} - \phi_{2}} \big( D\tilde{\zeta}_0 -D\tilde{\zeta}_1 \big)^2   \right. \\[3mm]
&& \left. \,\,\,\,\,\,   + \, e^{\phi_{1} + \phi_{2}} \Big( D\zeta^0 - D\zeta^1  + h \, ( D\tilde{\zeta}_0 -D\tilde{\zeta}_1  ) \Big)^2 \right. \\[3mm]
&& \left. \,\,\,\,\,\,  + \,   e^{\phi_{1} + \phi_{2}} \Big( D\tilde{\zeta}_0 +D\tilde{\zeta}_1  - h \, ( D\zeta^0 + D\zeta^1  ) \Big)^2  \right]  \ ,
\end{array}
\end{equation}
where we have introduced the kinetic term notation $\,dX \, dY \equiv \partial_{\mu}X \, \partial^{\mu}Y\,$ and $\,DX \, DY \equiv D_{\mu}X \, D^{\mu}Y\,$ for two generic scalars $\,X\,$ and $\,Y$. The covariant derivatives in (\ref{scalar_kinetic_SO(2,4)}) include a gauge connection for the gauge group in (\ref{G_N=2}). They can be straightforwardly computed from (\ref{cov_der_SO(6,6)}) and take the form
\begin{equation}
\label{cov_derivatives_1}
D\phi_{2} = d\phi_{2} - 4 g \, A^{^{(\epsilon)}}  \, h
\hspace{4mm} , \hspace{4mm}
Dh = dh - 2 g \, A^{^{(\epsilon)}} \, \left( e^{-2 \phi_{2}} - h^2 \right)
\hspace{4mm} , \hspace{4mm}
Da = da - 2 g \, A^{^{(a)}} \ ,
\end{equation}
together with
\begin{equation}
\label{cov_derivatives_2}
\begin{array}{ccc}
D\zeta^{0} = d\zeta^{0} - g \, A^{^{(\epsilon)}} \, (\tilde{\zeta}_0+\tilde{\zeta}_1) - 2 g \, A^{^{(\gamma)}} \,  \tilde{\zeta}_0 
& , &
D\zeta^{1} =  d\zeta^{1} - g \, A^{^{(\epsilon)}} \, (\tilde{\zeta}_0+\tilde{\zeta}_1) + 2 g \, A^{^{(\gamma)}} \,  \tilde{\zeta}_1  \ ,   
\\[3mm]
D\tilde{\zeta}_0 = d\tilde{\zeta}_0 - g \, A^{^{(\epsilon)}} \, (\zeta^1 - \zeta^0) + 2 g \, A^{^{(\gamma)}} \,  \zeta^0  \, 
& , & 
D\tilde{\zeta}_1 = d\tilde{\zeta}_1 - g \, A^{^{(\epsilon)}} \, (\zeta^0 - \zeta^1) - 2 g \, A^{^{(\gamma)}} \,  \zeta^1 \ , 
\end{array}
\end{equation}
in terms of three linear combinations of vectors $\,A_{\mu}{}^{\alpha M}\,$ given by
\begin{equation}
\label{A_vectors}
\begin{array}{rcll}
A^{^{(\epsilon)}}_{\mu} & \equiv & \sqrt{2} \, c \, \frac{1-\tilde{\varphi}^2}{1+\tilde{\varphi}^2} \, A_{\mu}{}^{+3} + \frac{\sqrt{2}}{\sqrt{1+\tilde{\varphi}^2}} \left( A_{\mu}{}^{+\bar{3}} + c \, \tilde{\varphi} \, A_{\mu}{}^{-3} \right) & , \\[4mm]
A^{^{(a)}}_{\mu} & \equiv &  \sqrt{2} \, c \, \frac{1-\tilde{\varphi}^2}{1+\tilde{\varphi}^2} \, A_{\mu}{}^{-\bar{4}} + \frac{\sqrt{2}}{\sqrt{1+\tilde{\varphi}^2}} \left( A_{\mu}{}^{-4} + c \, \tilde{\varphi} \, A_{\mu}{}^{+\bar{4}} \right)      & ,  \\[4mm]
A^{^{(\gamma)}}_{\mu} & \equiv &   \frac{\sqrt{2}}{\sqrt{1+\tilde{\varphi}^2}} \left( A_{\mu}{}^{+3} + A_{\mu}{}^{- \bar{4}} \right)  \ .

\end{array}
\end{equation}
The vectors $\,A^{^{(\epsilon)}}_{\mu}\,$, $\,A^{^{(a)}}_{\mu}\,$ and $\,A^{^{(\gamma)}}_{\mu}\,$ in (\ref{A_vectors}) respectively gauge the factors $\,\mathbb{R}_{\epsilon}\,$ and $\,\mathbb{R}_{a}\,$ and $\,\textrm{U}(1)_\gamma\,$ in (\ref{G_N=2}). There is an additional vector field $\,A^{^{(\textrm{R})}}_{\mu}\,$ associated with the $\,\textrm{U}(1)_{\textrm{R}}\,$ generator in (\ref{t_U(1)}) under which all the scalars in this sector of the theory are invariant.

\subsection{Warming up: the $(\varphi,\chi)$-family of $\,\mathcal{N}=2\,$ AdS$_{4}$ solutions of \cite{Bobev:2021yya}}

Let us start by recovering the two-parameter $(\varphi,\chi)$-family of $\,\mathcal{N}=2\,$ AdS$_{4}$ solutions of the $\,[\textrm{SO}(1,1) \times \textrm{SO}(6)] \ltimes \mathbb{R}^{12}\,$ maximal supergravity put forward in \cite{Bobev:2021yya}. As explained in Section~\ref{sec:Deforming_ISO(3)xISO(3)}, we must first of all set $\,\tilde{\varphi}^2 = 1\,$ in order to make contact with the maximal theory. Then, the two-parameter $(\varphi,\chi)$-family of $\,\mathcal{N}=2\,$ AdS$_{4}$ solutions of \cite{Bobev:2021yya} is recovered within the $\textrm{U}(1)_\textrm{R}$-invariant sector of half-maximal supergravity as follows. The SK scalars are fixed to
\begin{equation}
\label{U(1)_solution_1_maximal}
z_{5} = z_{7} = \dfrac{\mp 1+i}{\sqrt{2}}
\hspace{8mm} \textrm{ and } \hspace{8mm}
z_{2}= i \,  c  \ ,
\end{equation}
where, as before, the $\,\mp\,$ sign in (\ref{U(1)_solution_1_maximal}) is correlated with the $\,\pm\,$ sign in (\ref{ET_N=4_N=3_vac}). The QK scalars are fixed to
\begin{equation}
\label{U(1)_solution_2_maximal}
\begin{array}{c}
h + i \, e^{-\phi_{2}} = i \, \dfrac{\sqrt{2}}{c \, (1+\varphi^2)}
\hspace{5mm} , \hspace{5mm}
e^{-\phi_{1}} = \dfrac{c}{\sqrt{2}} \ , \\[6mm]
a = 0
\hspace{5mm} , \hspace{5mm}
\zeta^{0} = \zeta^{1} = \dfrac{c \, \chi}{\sqrt{2}}
\hspace{5mm} , \hspace{5mm}
\tilde{\zeta}_{0}= \tilde{\zeta}_{1} = \dfrac{\varphi}{\sqrt{1+\varphi^2}} \ .
\end{array}
\end{equation}

\noindent This $\,\mathcal{N}=2\,$ family of AdS$_{4}$ solutions comes along with a vacuum energy
\begin{equation}
V_{0}=-3 g^2 c^{-1} \ ,   
\end{equation}
for any value of the moduli fields $\,(\varphi,\chi)\,$ thus identifying them with marginal deformations in the dual CFT$_{3}$'s. It is worth emphasising that (\ref{U(1)_solution_1_maximal})-(\ref{U(1)_solution_2_maximal}) provides an explicit realisation of the $(\varphi,\chi)$-family of $\,\mathcal{N}=2\,$ AdS$_{4}$ solutions of \cite{Bobev:2021yya} in a specific supergravity model.\footnote{This two-parameter $(\varphi,\chi)$-family of solutions was constructed in \cite{Bobev:2021yya} by applying a $\,\chi$-dependent $\,\textrm{E}_{7(7)}\,$ duality transformation to the one-parameter $\varphi$-family in (\ref{vacuum_N=8_N=2_vac}).}

\subsubsection*{Marginal deformation and $\mathfrak{osp}(2|4)\,$ superconformal multiplets}

The explicit computation of the normalised mass spectrum recovers the half-maximal ($\mathbb{Z}_{2}$-invariant) subset of multiplets within the maximal content of \cite{Bobev:2021yya}. This consists of five unprotected long multiplets
\begin{equation}
\label{long_multiplets_N=4_varphi_chi}
L\bar{L}[0]^{0}_{\Delta_1}
\hspace{5mm} ,  \hspace{5mm} 
L\bar{L}[\tfrac{1}{2}]^{0}_{\Delta_{\pm}}
\hspace{5mm} ,  \hspace{5mm} 
L\bar{L}[0]^0_{\tilde{\Delta}_\pm} 
\ ,
\end{equation}
with conformal dimensions\footnote{The conformal dimensions $\,(\Delta_{1},\Delta_{+},\Delta_{-},\tilde{\Delta}_{+},\tilde{\Delta}_{-})\,$ map into $\,(\beta_{3},\beta_{5},\beta_{4},\beta_{1},\beta_{2})\,$ in eq.$(4.5)$ of \cite{Bobev:2021yya} upon the identification $\,\chi = \sqrt{1+\tilde{\varphi}^2}\, \chi_{\tiny{\cite{Bobev:2021yya}}}$ with $\,\tilde{\varphi}=1\,$.}

\begin{equation}
\begin{array}{cll}
\Delta_1 &=& \frac{1}{2} +  \frac{1}{2} \sqrt{\frac{17+33 \varphi^2}{1+\varphi ^2}}  \ , \\[6mm]
\Delta_{\pm} &=&  \frac{1}{2}+ \frac{\sqrt{\left(2+\varphi^2\right)^2 + \chi ^2} \, \pm \, \varphi}{\sqrt{2\left(1+\varphi^2\right)}} \ , \\[6mm]
\tilde{\Delta}_\pm &=& \frac{1}{2} +  \frac{1}{2}\sqrt{\frac{9+4 \varphi^4+5 \varphi^2+ 4 \chi^2 \pm 4 (\varphi ^4+\varphi ^2+ \chi ^2-2)}{\varphi ^2+1}}\ .
\end{array}
\end{equation}
There are also the short and semi-short protected multiplets given in (\ref{short_maximal_theory}), namely,
\begin{equation}
\label{short_N=4_theory_bobev}
A_1\bar{A}_1[1]_2^0
\hspace{5mm} , \hspace{5mm}
L\bar{B}_1[0]_2^2
\hspace{5mm} , \hspace{5mm}
B_1\bar{L}[0]_2^{-2} \ ,
\end{equation}
where $\,A_1\bar{A}_1[1]^{0}_2\,$ is identified with the stress-energy tensor multiplet of the $\,\mathcal{N}=2\,$ CFT$_{3}$'s.

\subsection{A $\,(\varphi,\chi \,;\tilde{\varphi})$-family of $\,\mathcal{N}=2\,$ AdS$_{4}$ solutions}

The two-parameter family of $\,\mathcal{N}=2\,$ AdS$_{4}$ solutions in (\ref{U(1)_solution_1_maximal})-(\ref{U(1)_solution_2_maximal}) can be generalised to arbitrary values of the deformation parameter $\,\tilde{\varphi}\,$. The SK scalars are given by
\begin{equation}
\label{U(1)_solution_1}
z_{5} = z_{7} = \dfrac{\mp\tilde{\varphi}+i}{\sqrt{1+\tilde{\varphi}^2}}
\hspace{8mm} \textrm{ and } \hspace{8mm}
z_{2}= i \,  c  \ ,
\end{equation}
whereas the QK scalars take the form
\begin{equation}
\label{U(1)_solution_2}
\begin{array}{c}
h + i \, e^{-\phi_{2}} = i \, \dfrac{\sqrt{1+\tilde{\varphi}^2}}{c \, (1+\varphi^2)}
\hspace{5mm} , \hspace{5mm}
e^{-\phi_{1}} = \dfrac{c}{\sqrt{1+\tilde{\varphi}^2}} \ , \\[6mm]
a = 0
\hspace{5mm} , \hspace{5mm}
\zeta^{0} = \zeta^{1} = \dfrac{c \, \chi}{\sqrt{1+\tilde{\varphi}^2}}
\hspace{5mm} , \hspace{5mm}
\tilde{\zeta}_{0}= \tilde{\zeta}_{1} = \dfrac{\varphi}{\sqrt{1+\varphi^2}} \ .
\end{array}
\end{equation}

\noindent The vacuum energy at this $\,\mathcal{N}=2\,$ family of AdS$_{4}$ solutions is still given by
\begin{equation}
V_{0}=-3 g^2 c^{-1} \ ,   
\end{equation}
for any value of the deformation parameter $\,\tilde{\varphi}\,$ as well as of the moduli fields $\,(\varphi,\chi)\,$ dual to marginal deformations. It is worth highlighting that, at any value of $\,\tilde{\varphi}$, turning on $\,(\varphi,\chi)\,$ activates the hypermultiplet scalars $\,(\zeta^{0},\zeta^{1} ; \tilde{\zeta}_{0},\tilde{\zeta}_{1})\,$ spanning the Heisenberg fiber of the QK geometry. Activating these scalars automatically breaks the compact $\,\textrm{U}(1)_{\gamma}\,$ factor of the gauge group (\ref{G_N=2}), as it can be seen from (\ref{cov_derivatives_2}).

\subsection{$\mathfrak{osp}(2|4)\,$ superconformal multiplets}

The half-maximal supergravity spectrum \textit{at generic} $\,\tilde{\varphi}\,$ of the $\,\mathcal{N}=2\,$ $(\varphi,\chi)$-family of AdS$_{4}$ solutions  can be arranged into multiplets of the $\,\mathfrak{osp}(2|4)\,$ superconformal symmetry of the would-be dual $\,\mathcal{N}=2\,$ CFT$_{3}$'s. The spectrum contains five unprotected long multiplets
\begin{equation}
\label{long_multiplets_general_U(1)}
L\bar{L}[0]^0_{\Delta_1} 
\hspace{5mm} ,  \hspace{5mm} 
L\bar{L}[\tfrac{1}{2}]^{0}_{\Delta_\pm}
\hspace{5mm} ,  \hspace{5mm} 
L\bar{L}[0]^{0}_{\tilde{\Delta}_\pm}  \ ,
\end{equation}
with conformal dimensions given by
\begin{equation}
\label{Deltas_general_U(1)_tilde_varphi}
\begin{array}{cll}
\Delta_1 &=& \frac{1}{2} +  \frac{1}{2} \sqrt{\frac{9+25 \tilde{\varphi}^2+\varphi^2 \left(17+49 \tilde{\varphi}^2\right)}{\left(1+\varphi^2\right) \left(1+\tilde{\varphi}^2\right)}} \ , \\[6mm]
\Delta_{\pm} &=&  \frac{1}{2}+ \frac{\sqrt{\left(1+\varphi^2+\tilde{\varphi}^2\right)^2 +  \chi ^2} \, \pm \, \varphi  \tilde{\varphi}}{\sqrt{\left(1+\varphi^2\right) \left(1+\tilde{\varphi}^2\right)}} \ , \\[6mm]
\tilde{\Delta}_\pm &=& \scriptsize{\frac{1}{2}+\frac{1}{2}\sqrt{\frac{5+8 \varphi ^4+\varphi ^2 \left(\tilde{\varphi}^2+9\right)+8 \tilde{\varphi}^4+5 \tilde{\varphi}^2+8 \chi ^2 \pm 4 \sqrt{\Theta}}{\left(\varphi ^2+1\right) \left(\tilde{\varphi}^2+1\right)}}} \ ,
\end{array}
\end{equation}
where
\begin{equation}
\begin{array}{lll}        
\Theta & = & 4 \varphi ^8 + 8 \varphi ^6 - 4 \varphi ^4 \left(\tilde{\varphi}^4+3 \tilde{\varphi}^2-1-2 \chi ^2\right) - 4 \varphi ^2 \left(\tilde{\varphi}^4+3 \tilde{\varphi}^2-2 \chi ^2\right)\\[2mm]
& & \left(\tilde{\varphi}^2 \left(1+2 \tilde{\varphi}^2\right)+1-2 \chi ^2\right)^2 + 4 \left(\tilde{\varphi}^2-1\right)^2 \chi ^2 .
\end{array}
\end{equation}
In addition, there are one short and two semi-short protected multiplets with integer conformal dimension $\,\Delta=2\,$. These are the same as in (\ref{short_maximal_theory}), namely,
\begin{equation}
\label{short&semi-short_multiplets_general_Family_I}
A_1\bar{A}_1[1]^{0}_{2} 
\hspace{5mm} ,  \hspace{5mm} 
L\bar{B}_1[0]^{2}_{2} 
\hspace{5mm} ,  \hspace{5mm} 
B_1\bar{L}[0]^{-2}_{2}   \ ,
\end{equation}
where $\,A_1\bar{A}_1[1]^{0}_2\,$ is the stress-energy tensor multiplet of the dual $\,\mathcal{N}=2\,$ CFT$_{3}$'s. The two moduli fields $\,\varphi\,$ and $\,\chi\,$ in (\ref{U(1)_solution_2}) belong to the semi-short multiplets $\,L\bar{B}_1[0]^{2}_{2}\,$ and $\,B_1\bar{L}[0]^{-2}_{2}$.

\subsection{Special loci}

The computation of the four gravitino masses in the half-maximal theory yields
\begin{equation}
\label{Gravitino_masses_N4_general}
m \, L  =  1 \,\,\,\, (\times 2) 
\hspace{5mm} , \hspace{5mm}
\frac{\sqrt{(1+\varphi^2+\tilde{\varphi}^2)^2 + \chi^2} \pm \varphi \, \tilde{\varphi}}{\sqrt{\left( 1+\varphi^2 \right) \left( 1+\tilde{\varphi}^2 \right)}}  \ ,
\end{equation}
which consistently reduces to (\ref{Gravitino_masses_N4}) when setting $\,\chi=0\,$. Notice that, even when turning on $\,\chi \neq 0\,$, the marginal deformation $\,\varphi\,$ and the embedding tensor parameter $\,\tilde{\varphi}\,$ continue entering the gravitino masses (\ref{Gravitino_masses_N4_general}) in a symmetric fashion. A quick inspection of (\ref{Gravitino_masses_N4_general}) shows that supersymmetry enhancement to $\,\mathcal{N}>2\,$ is no longer possible whenever $\,\chi \neq 0$. Still we will look at two special cases. The first case is $\,\tilde{\varphi}=0\,$ which accounts for the $\,\mathcal{N}=2\,$ marginal deformations of the $\,\mathcal{N}=4\,$ exotic CFT$_{3}$. The second one is $\,\varphi=0\,$ which describes the effect of the modulus $\,\chi\,$ in a genuine half-maximal supergravity at generic $\,\tilde{\varphi}$.

\subsubsection{$(\varphi,\chi)$-deformations of the $\,\mathcal{N}=4\,$ exotic CFT$_{3}$}
\label{sec:deformation_exotic}

Setting $\,\tilde{\varphi}=0\,$ in the general expressions of the previous section one is left with the $\,\mathcal{N}=2\,$ $\,(\varphi,\chi)$ marginal deformations of the exotic $\,\mathcal{N}=4\,$ CFT$_{3}$. More concretely, we find in this case five unprotected long multiplets
\begin{equation}
L\bar{L}[0]^{0}_{\Delta_1}
\hspace{5mm} ,  \hspace{5mm} 
L\bar{L}[\tfrac{1}{2}]^{0}_{\Delta} \,\,\,\, (\times 2) 
\hspace{5mm} ,  \hspace{5mm} 
L\bar{L}[0]^0_{\tilde{\Delta}_\pm} 
\ ,
\end{equation}
with conformal dimensions
\begin{equation}
\begin{array}{cll}
\Delta_1 &=& \frac{1}{2} + \frac{1}{2} \sqrt{\frac{9 + 17\varphi^2}{1+\varphi^2}} \ , \\[6mm]
\Delta &=&  \frac{1}{2} +\frac{\sqrt{\left(1+\varphi^2\right)^2+\chi^2}}{\sqrt{1+\varphi^2}}\ , \\[6mm]
\tilde{\Delta}_\pm &=& \frac{1}{2} + \frac{1}{2} \sqrt{\frac{5 + 8 \chi ^2 +8 \varphi ^4+9 \varphi ^2\pm4 \sqrt{1+4 \left(\varphi ^4+\varphi ^2+\chi ^2\right)^2}}{1+\varphi^2}}\ .
\end{array}
\end{equation}
Notice that, unlike for the $\,\mathcal{N}=4 \,\&\, \textrm{SO}(4)_{\textrm{R}}\,$ S-fold in (\ref{long_multiplets_N=4_varphi_chi}), the multiplets $\,L\bar{L}[\tfrac{1}{2}]^{0}_{\Delta_\pm}\,$ get degenerated in this case. In addition, there are also the short and semi-short protected multiplets given in (\ref{short&semi-short_multiplets_general_Family_I}). The long multiplets $\,L\bar{L}[\tfrac{1}{2}]^{0}_{\Delta}\,$ and $\,L\bar{L}[0]^{0}_{\tilde{\Delta}_{-}}\,$ hit the unitarity bound at the special value $\,\varphi = \chi= 0\,$ and split as
\begin{equation}
\label{splitting_varphitilde=0_family_I}
\begin{array}{lcl}
L\bar{L}[\tfrac{1}{2}]^{0}_{\frac{3}{2}} & \rightarrow & A_1\bar{A}_1[\tfrac{1}{2}]_{\frac{3}{2}}^0 \,\, \oplus \,\, A_2\bar{L}[0]_2^{-1} \,\, \oplus \,\, L\bar{A}_2[0]_2^{1}   \\[4mm]
L\bar{L}[0]^{0}_{1} & \rightarrow &  A_2\bar{A}_2[0]_1^0 \,\, \oplus \,\, B_1\bar{L}[0]_2^{-2} \,\, \oplus \,\, L\bar{B}_1[0]_2^{2}
\end{array}
\end{equation}
recovering the undeformed exotic $\,\mathcal{N}=4\,$ CFT$_{3}$.

\subsubsection{$(\chi\,;\tilde{\varphi})$-family of $\,\mathcal{N}=2\,$ AdS$_{4}$ solutions}

In Section~\ref{sec:family_1} we characterised the $\,(\varphi\,;\tilde{\varphi})$-family of $\,\mathcal{N}=2\,$ CFT$_{3}$'s at $\,\chi=0\,$. Let us now take the complementary case $\,\varphi=0\,$ and characterise the $\,(\chi\,;\tilde{\varphi})$-family of $\,\mathcal{N}=2\,$ CFT$_{3}$'s. Since the axion-like deformations are by now well understood geometrically for the S-fold backgrounds at $\,\tilde{\varphi}=1\,$ \cite{Guarino:2021kyp}, it would be interesting to investigate whether a higher-dimensional geometric interpretation at arbitrary values of $\,\tilde{\varphi}\,$ is still be possible.

Setting $\,\varphi=0\,$, the half-maximal spectrum of dual operators contains the five unprotected long multiplets
\begin{equation}
L\bar{L}[0]_{\Delta_1}^0
\hspace{5mm} , \hspace{5mm}
L\bar{L}[\tfrac{1}{2}]_{\Delta}^0 \,\,\,\, (\times 2) 
\hspace{5mm} , \hspace{5mm}
L\bar{L}[0]_{\tilde{\Delta}_{\pm}}^0   \ ,
\end{equation}
this time with conformal dimensions given by
\begin{equation}
\label{tildevarphi-chi-family}
\begin{array}{ccl}
\Delta_{1} &=& \frac{1}{2} + \frac{1}{2} \sqrt{\frac{9+25 \tilde{\varphi}^2}{1+\tilde{\varphi}^2}}  \ , \\[4mm] 
\Delta &=& \frac{1}{2} + \frac{\sqrt{\left(1+\tilde{\varphi}^2\right)^2 + \chi^2}}{\sqrt{1+\tilde{\varphi}^2}} \ , \\[4mm] 
\tilde{\Delta}_{\pm} &=& \frac{1}{2}  + \frac{1}{2} \sqrt{\frac{8 \tilde{\varphi}^4+5 \tilde{\varphi}^2+8 \chi ^2+5\pm 4 \sqrt{\Theta_{\varphi\to 0}}}{1+\tilde{\varphi}^2}} \ , 
\end{array}
\end{equation}
with
\begin{equation}
    \Theta_{\varphi\to 0} = \left(\tilde{\varphi}^2 \left(1+2 \tilde{\varphi}^2\right)+1-2 \chi ^2\right)^2 + 4 \left(\tilde{\varphi}^2-1\right)^2 \chi ^2 .
\end{equation}
There are also the short and semi-short protected multiplets given in (\ref{short&semi-short_multiplets_general_Family_I}). As a check of consistency, the $\,\chi$-family of $\,\mathcal{N}=2\,$ S-folds in \cite{Guarino:2020gfe} is recovered at $\,\tilde{\varphi} = 1$. We also recover the results of Section~\ref{sec:chi_def_exotic_N=4} upon setting $\,\tilde{\varphi}=0\,$ together with $\chi_1 = - \chi_3= \chi\,$ and $\,\chi_2= 0\,$ (up to a $\,\textrm{U}(1)_\gamma\,$ transformation (see eq.(\ref{gauge_gamma})) of angle $\,\gamma = - \frac{\pi}{4}$).

\subsection{On the conformal manifold of $\,\mathcal{N}=2\,$ CFT$_{3}$'s}

Given the supergravity model in Section~\ref{sec:N=2_model}, which includes vector fields and gaugings of scalar isometries, the holographic Zamolodchikov metric on a conformal manifold of CFT$_{3}$'s cannot be obtained simply by direct substitution of the AdS$_{4}$ solution (\ref{U(1)_solution_1})-(\ref{U(1)_solution_2}) into the scalar kinetic terms (\ref{scalar_kinetic_SO(2,4)}). The reason being that a solution like (\ref{U(1)_solution_1})-(\ref{U(1)_solution_2}) can be brought to a different, but physically equivalent, form upon a gauge transformation.

The infinitesimal $\,\textrm{U}(1)_\gamma \times \mathbb{R}_a \times \mathbb{R}_\epsilon\,$ gauge transformations entering the covariant derivatives in (\ref{cov_derivatives_2}) -- recall that all the fields within this sector are invariant under $\,\textrm{U}(1)_\textrm{R}$ -- can be integrated to finite transformations. To describe such finite transformations we will introduce three complex fields
\begin{equation}
z = h + i \,e^{-\phi_2}
\hspace{5mm} , \hspace{5mm}
\psi_0 = \zeta_0 + i\, \tilde{\zeta}_0
\hspace{5mm} , \hspace{5mm}
\psi_1 = \tilde{\zeta}_1 + i\, \zeta_1 \ ,
\end{equation}
in terms of which the compact $\,\textrm{U}(1)_\gamma\,$ transformation acts as 
\begin{equation}
\label{gauge_gamma}
\psi_0 \rightarrow e^{i\,\gamma} \, \psi_0 
\hspace{1cm} \text{ and } \hspace{1cm} 
\psi_1 \rightarrow e^{i\,\gamma} \, \psi_1 \ ,
\end{equation}
the non-compact $\,\mathbb{R}_a\,$ acts as a shift
\begin{equation}
\label{gauge_a}
a \rightarrow a + c_{a} \ ,
\end{equation}
and $\,\mathbb{R}_\epsilon\,$ acts as a fractional linear transformation  
\begin{equation}
\label{gauge_epsilon}
z \rightarrow \frac{z}{\epsilon \,z+1} \hspace{10mm}\text{and}\hspace{10mm} 
\begin{pmatrix}
\psi_0 \\
\psi_1
\end{pmatrix} 
\rightarrow 
\begin{pmatrix} 
1- i\, \frac{\epsilon}{2} & \frac{\epsilon}{2} \\
\frac{\epsilon}{2} & 1 + i\,\frac{\epsilon}{2}
\end{pmatrix} \, 
\begin{pmatrix}
\psi_0\\
\psi_1
\end{pmatrix} \ .
\end{equation}
As a result, the gauge-fixed solution in (\ref{U(1)_solution_1})-(\ref{U(1)_solution_2}) can be gauge-released by acting on it with (\ref{gauge_gamma})-(\ref{gauge_epsilon}). This action introduces three additional (yet unphysical) parameters $\,(\gamma,c_{a},\epsilon)\,$ in the solution (\ref{U(1)_solution_1})-(\ref{U(1)_solution_2}), and the naive pull-back of the metric (\ref{scalar_kinetic_SO(2,4)}) on different gauge-fixed solutions will depend on the choice of gauge. For example, choosing $\,c_{a} = f(\varphi,\chi)\,$ in (\ref{gauge_a}) and performing the pull-back of the metric (\ref{scalar_kinetic_SO(2,4)}), one encounters different Zamolodchikov metrics for different choices of the function $\,f(\varphi,\chi)\,$. Of course, the catch is that we are considering solutions which are gauge-equivalent, and therefore physically equivalent, as different. In order to perform the gauge-fixing properly, one should not pick up a subspace within the gauge-released space of solutions and perform the naive pull-back of the metric onto it. Instead, one must study the quotient of the gauge-released space of solutions by the gauge group. This quotient space is the one being dual to a conformal manifold with a uniquely defined Zamolodchikov metric.

For the sake of concreteness, let us particularise to our specific gauge-released or ambient scalar geometry as described by the kinetic terms in (\ref{scalar_kinetic_SO(2,4)}). We first need to identify three independent one-forms on the scalar geometry which are to be declared as ``pure gauge" or unphysical, and then quotient the scalar geometry by them. How to identify such three one-forms in field space is a physical question. And the answer to that question comes from the vector sector which, despite being set to zero at the supergravity solution, still provides equations of motion that must hold. In short, we must quotient the geometry by the one-form currents $\,J^{^{(\gamma)}}$, $\,J^{^{(a)}}$ and $\,J^{^{(\epsilon)}}$ acting as sources for the vector fields that have been set to zero at the supergravity solution. This implies that we should first put those one-forms to zero in the kinetic terms (\ref{scalar_kinetic_SO(2,4)}) before reading off the Zamolodchikov metric by performing the pull-back of the ambient metric on any gauge-fixed subspace of solutions.

Let us illustrate the procedure described above by looking at the gauge-fixing of the $\,\mathbb{R}_{a}\,$ shift symmetry in (\ref{gauge_a}) spanned by the vector field $\,A^{^{(a)}}$. The scalar $\,a\,$ plays the role of a St\"uckelberg field for the massive vector $\,A^{^{(a)}}$ -- recall that $\,Da = da + 2 \, g \, A^{^{(a)}}$ in (\ref{cov_derivatives_1}) -- and the associated current computed from (\ref{scalar_kinetic_SO(2,4)}) reads
\begin{equation}
\label{EOM_vector}
J^{^{(a)}} \equiv \, g \, e^{2\phi_{1} } * \left[ Da + \tfrac12 \big( \zeta^0  D\tilde{\zeta}_0 + \zeta^1  D\tilde{\zeta}_1 - \tilde{\zeta}_0 D\zeta^0 - \tilde{\zeta}_1 D\zeta^1  \big) \right] \ .
\end{equation}
Quotienting the scalar geometry by this (field space) one-form implies that (\ref{EOM_vector}) must be set to zero identically, \textit{i.e.} $J^{^{(a)}}=0\,$, when evaluating any quantity at a gauge-fixed solution like (\ref{U(1)_solution_1})-(\ref{U(1)_solution_2}). In particular, its contribution to the third line in the kinetic terms (\ref{scalar_kinetic_SO(2,4)}) must be removed before reading off the Zamolodchikov metric from it. Proceeding similarly with the contributions coming from the remaining one-form currents $\,J^{^{(\epsilon)}}\,$ and $\,J^{^{(\gamma)}}\,$ acting as sources for $\,A^{^{(\epsilon)}}\,$ and $\,A^{^{(\gamma)}}$, the resulting Zamolodchikov metric becomes independent of the deformation parameter $\,\tilde{\varphi}\,$ and reads
\begin{equation}
\label{CM_metric_gauge_fixed}
ds_{\textrm{CM}}^2 = \dfrac{1+2 \varphi^2}{2(1+\varphi^2)^2} \big( d\varphi^2 + (1+\varphi^2) \, d\chi^2 \big) \ .
\end{equation}
This metric in the conformal manifold of $\,\mathcal{N}=2\,$ CFT$_{3}$'s at generic $\,\tilde{\varphi}\,$ agrees with that of \cite{Bobev:2021yya} upon the identification $\,\chi = \sqrt{1+\tilde{\varphi}^2}\, \chi_{\scriptsize{\cite{Bobev:2021yya}}}$ with $\,\tilde{\varphi}=1\,$.

The moduli $\,(\varphi,\chi)\,$ belong to the hypermultiplet sector in the AdS$_{4}$ solution (\ref{U(1)_solution_1})-(\ref{U(1)_solution_2}). Recalling that, for AdS$_{4}$ solutions preserving $\,\mathcal{N}=2\,$, the hypermultiplet moduli space must be a K\"ahler submanifold of the quaternionic K\"ahler geometry \cite{deAlwis:2013jaa}, the conformal manifold in (\ref{CM_metric_gauge_fixed}) must be K\"ahler. The K\"ahlericity of the metric (\ref{CM_metric_gauge_fixed}) can be checked as follows. Let us first introduce a set of so-called isothermal coordinates for which the metric is conformal to the Euclidean metric. These are given by $\,x=\chi\,$ and $\,y=\textrm{arcsinh}\varphi\,$ so that the Zamolodchikov metric in (\ref{CM_metric_gauge_fixed}) is brought to the form
\begin{equation}
ds_{\textrm{CM}}^2 =\tfrac{1}{2} \, \Omega(y)^2 \left( dx^2 + dy^2 \right)
\hspace{8mm} \textrm{ with } \hspace{8mm}
\Omega(y)^2 \equiv 1+\tanh^2y \ .
\end{equation}
Introducing the complex coordinate $\,z=x+iy\,$ one arrives at  
\begin{equation}
ds_{\textrm{CM}}^2 = g_{z\bar{z}} \, dz \, d\bar{z}
\hspace{8mm} \textrm{ with } \hspace{8mm}
g_{z\bar{z}} = \tfrac{1}{2} \, \left( 1 + \tanh^2\left[\tfrac{-i(z-\bar{z})}{2}\right]\right) \ ,
\end{equation}
where $\,g_{z\bar{z}}=\frac{\partial^2 K}{\partial z \partial \bar{z}}\,$ can be derived from the real K\"ahler potential
\begin{equation}
K(z,\bar{z}) = |z|^2 - \log\left[\cosh^2\left(\tfrac{-i(z-\bar{z})}{2}\right) \right] \ .
\end{equation}

\subsubsection*{More on gauge-fixing and Zamolodchikov metric}

The scalar manifold $\,\mathcal{M}_{\text{scal}}\,$ in (\ref{M_scalar}) of half-maximal supergravity is endowed with a canonical Riemannian metric $\,g$, prior to any gauge fixing, and a left action of a gauge group $\,\textrm{G}\,$, \textit{e.g.}, $\,\textrm{G}=\textrm{ISO}(3)_{1} \times \textrm{ISO}(3)_{2}\,$ in our case. In general, the action of $\,\textrm{G}\,$ on $\,\mathcal{M}_{\text{scal}}\,$ is not free -- there are fixed loci under the action of the compact part of $\,\textrm{G}\,$ -- and thus the quotient space $\,\textrm{G}\backslash\mathcal{M}_{\text{scal}}\,$ is not a manifold. In supergravity, the action of $\,\textrm{G}\,$ on $\,\mathcal{M}_{\text{scal}}\,$ is well-behaved and we can chop the fixed loci out of $\,\mathcal{M}_{\text{scal}}\,$ to define a new manifold $\,\tilde{M}_{\text{scal}}\,$ on which $\,\textrm{G}\,$ has a free action. 

In order to establish a connection with the $\,\mathcal{N}=2\,$ supergravity model of Section~\ref{sec:N=2_model}, we will focus on the $\,\textrm{U}(1)_{\textrm{R}}$-invariant sector of half-maximal supergravity for which $\,\textrm{G} \rightarrow {\textrm{G}_{\mathcal{N}=2} =  \textrm{U}(1)_\textrm{R} \times \textrm{U}(1)_\gamma \times \mathbb{R}_a \times \mathbb{R}_\epsilon}\,$ and $\,\mathcal{M}_{\text{scal}}\,$ in (\ref{M_scalar})  reduces to the one in (\ref{scalar_geometry_N2}). From (\ref{cov_derivatives_2}), the $\,\textrm{U}(1)_\textrm{R} \times \textrm{U}(1)_\gamma\,$ compact part of $\,\textrm{G}_{\mathcal{N}=2}\,$ leaves invariant the scalar locus defined by the condition $\,\tilde{\zeta}_0=\tilde{\zeta}_1=\zeta^0=\zeta^1=0\,$.\footnote{In the gauge-fixed supergravity solution of (\ref{U(1)_solution_1})-(\ref{U(1)_solution_2}), this locus corresponds to setting to zero the marginal deformations, \textit{i.e.}, $\,\varphi=\chi=0\,$.} Starting from the gauge-released solution extending the gauged-fixed one in (\ref{U(1)_solution_1})-(\ref{U(1)_solution_2}) with three parameters $\,(\gamma,c_{a},\epsilon)\,$ (see discussion below (\ref{gauge_epsilon})) and removing the fixed locus under $\,\textrm{U}(1)_\gamma$, we can finally define a manifold of supergravity solutions $\,\mathcal{S} \subset \tilde{\mathcal{M}}_{\text{scal}}\,$ on which $\,\textrm{G}_{\mathcal{N}=2}\,$ acts freely. This gives a structure of principal bundle
\begin{equation}
\pi : \mathcal{S} \rightarrow \textrm{G}_{\mathcal{N}=2} \backslash \mathcal{S}  \ .  
\end{equation}
The quotient space $\,\textrm{G}_{\mathcal{N}=2} \backslash \mathcal{S}\,$ is the object dual to the $\,\mathcal{N}=2\,$ conformal manifold of CFT$_{3}$'s, and it is on this quotient space that we must define metric $\,g_{\textrm{CM}}\,$ dual to the Zamolodchikov metric in (\ref{CM_metric_gauge_fixed}).

We can always decompose the tangent space of $\,\mathcal{S}\,$ as $\,T\mathcal{S} = V\mathcal{S} \oplus H\mathcal{S}\,$ where $\,V\mathcal{S} = \text{ker}\,\pi_*\,$ and $\,H\mathcal{S} = [V\mathcal{S}]^\perp\,$ is the orthogonal complement of $\,V\mathcal{S} \,$ with respect to the metric $\,g$. The tangent space $\,T\mathcal{S}\,$ should then be understood as the space of \textit{all} small deformations: $\,H\mathcal{S}\,$ corresponding to \textit{physical} deformations and $\,V\mathcal{S}\,$ being the space of \textit{unphysical} deformations. The latter are exactly the infinitesimal gauge transformations, and a projector $\,\text{Pr}_{H\mathcal{S}}\,$ onto $\,H\mathcal{S}\,$ can be defined that projects them away. Finally, for any $\,x \in \textrm{G}_{\mathcal{N}=2} \backslash \mathcal{S}\,$ there is a $\,p \in \mathcal{S}\,$ such that $\,\pi(p) = x$. Then, for any pair of vectors $\,(v,\,v')\,$ in $\,T(\textrm{G}_{\mathcal{N}=2}\backslash \mathcal{S})\,$ we can choose vectors $\,(w,\,w')\,$ in $\,T\mathcal{S}\,$ such that $\, \pi_*w = v\,$ and $\,\pi_* w' = v'$. In this manner the Zamolodchikov metric can be defined as the map
\begin{equation}
g_{\textrm{CM}} : T(\textrm{G}_{\mathcal{N}=2}\backslash \mathcal{S})^{\otimes 2} \rightarrow \mathbb{R} : v\otimes v' \rightarrow g_{\textrm{CM}}(v,\,v') = g\left( \, \text{Pr}_{H\mathcal{S}} w,\,\text{Pr}_{H\mathcal{S}}w' \,\right)\,.
\end{equation}
This is a well defined metric on $\,\textrm{G}_{\mathcal{N}=2}\backslash \mathcal{S}\,$ and it is equivalent to the prescription we have given in our example. Importantly, it does not depend on the choice of gauge fixing nor the invariant subsector used to find the supergravity solution within the full theory.

\section{Final remarks}
\label{sec:conclusions}

In the present paper we have initiated a holographic study of new CFT$_{3}$'s with $\,\mathcal{N}=2,3,4\,$ supersymmetry using an effective four-dimensional gauged supergravity approach. The rich structure of multi-parametric families of supersymmetric AdS$_{4}$ solutions we have just started to identify in the half-maximal gauged supergravities with $\,\textrm{ISO}(3)_{1} \times \textrm{ISO}(3)_{2}\,$ gaugings raises some immediate questions.

Perhaps the most obvious question is whether or not the $\,\tilde{\varphi}$-family of $\,\textrm{ISO}(3)_{1} \times \textrm{ISO}(3)_{2}\,$ gaugings of half-maximal supergravity we have presented in Sections \ref{sec:Deforming_ISO(3)xISO(3)} and \ref{sec:algebra_ISO(3)xISO(3)} (and, more ambitiously, its generalisation in Appendix~\ref{app:ISO(3)xISO(3)_general_gauging}) describes consistent truncations of ten- or eleven-dimensional supergravity down to four dimensions. In this respect, since turning on the embedding parameter $\,\tilde{\varphi}\,$ (\textit{i.e.} $\,\tilde{\varphi} \neq \pm 1\,$ in our parameterisation) only affects the embedding of the non-compact translational generators in the gauge algebra (see \textit{e.g.} (\ref{algebra_ISO3_1_family_I})-(\ref{algebra_ISO3_2_family_I})), the family of $\,\mathcal{N}=2\,$ AdS$_{4}$ solutions we have found may stand a chance of being upliftable to new (possibly only locally geometric) type II or M-theory backgrounds. However, it could still happen -- as for the $\,\omega\,$ deformation of the $\textrm{SO}(8)$-gauged supergravity \cite{Dall'Agata:2012bb} -- that only very specific values of $\,\tilde{\varphi}\,$ enjoy a higher-dimensional interpretation, the natural ones being $\,\tilde{\varphi} = \pm 1\,\textrm{ and } \, 0\,$. The case $\,\tilde{\varphi} = \pm 1\,$ is by now known to uplift to type IIB S-fold backgrounds. Examples are the $\,\mathcal{N}=4\,$ S-fold of \cite{Inverso:2016eet} and its marginal deformations \cite{Guarino:2020gfe,Bobev:2021yya,Guarino:2021hrc}. The case $\,\tilde{\varphi}=0\,$ remains to be understood. But it would certainly be disappointing if a supergravity solution like the exotic AdS$_{4}$ vacuum of \cite{Dibitetto:2011gm} with such a (conjectured but) highly symmetric $\,\mathcal{N}=4\,$ CFT$_{3}$ dual -- together with its marginal deformations presented in Sections \ref{sec:chi_def_exotic_N=4} and \ref{sec:deformation_exotic} -- ended up being in the Swampland.

Let us further comment on the exactly marginal deformation of the exotic $\,\mathcal{N}=4\,$ CFT$_{3}$ dual to the modulus $\,\chi\,$ in (\ref{U(1)_solution_2}) preserving $\,\mathcal{N}=2\,$ supersymmetry. As discussed in Section~\ref{sec:exotic_chi's}, turning on $\,\chi\,$ breaks the original $\,\textrm{SO(4)}_\textrm{R}\,$ symmetry down to a $\,\textrm{U}(1)_{\textrm{R}}\,$ factor within the Cartan subgroup $\,\textrm{U}(1)_{\textrm{R}} \times \textrm{U}(1)_{\textrm{F}} \subset \textrm{SO}(4)_{\textrm{R}}$. As a result, the whole Cartan subgroup is not preserved and a geometric interpretation of the modulus $\,\chi\,$ along the lines of the axion-like deformations of \mbox{S-folds} seems a priori unplausible within the context of half-maximal supergravity. However, the situation is more subtle: when setting $\,\tilde{\varphi}=\pm 1\,$, the embedding of this solution into maximal supergravity provides an additional flavour current multiplet $\,A_2\bar{A}_2[0]_1^0\,$ that accounts for the $\,\textrm{U}(1)_{\textrm{F}}\,$ symmetry \cite{Bobev:2021yya}. In this case, the $\,\textrm{U}(1)_{\textrm{F}}\subset \textrm{SO}(4)_{\textrm{R}}\,$ is actually not broken but projected out by the $\,\mathbb{Z}_{2}\,$ symmetry truncating maximal to half-maximal supergravity. Therefore, if a ten- or eleven-dimensional uplift exists for the AdS$_{4}$ solution (\ref{U(1)_solution_1})-(\ref{U(1)_solution_2}) at $\,\tilde{\varphi} = 0\,$ (or more generically at $\,\tilde{\varphi} \neq \pm 1\,$), it is still possible that some symmetries have been truncated away in the half-maximal effective description. Needless to say, an uplift (if any) to ten or eleven dimensions is required in order to settle this question.

The effective four-dimensional gauged supergravity approach adopted in this work provides us with some guidance in order to guess what the potential higher-dimensional realisation of the AdS$_{4}$ solutions could be. Remarkably, amongst the extra quadratic constraints in (\ref{QC_Extra}), only those in the $\,(\mathbf{1},\mathbf{462'})\,$ irrep of $\,\textrm{SL}(2) \times \textrm{SO}(6,6)\,$ are violated in the AdS$_{4}$ solutions with $\,\tilde{\varphi} \neq \pm 1$.\footnote{The extra quadratic constraints in (\ref{QC_Extra}) living in the $\,(\mathbf{3},\mathbf{1})\,$ irrep are not violated in our half-maximal supergravity models. These constraints appeared as the four-dimensional incarnation of the $\,\textrm{SL}(2)\,$ extension \cite{Ciceri:2016hup} of the so-called \textit{section constraint} in double field theory \cite{Hull:2009mi}.} Assuming (as for the S-folds) a type IIB origin, and since in a type IIB duality frame the $\,\textrm{SL}(2)\,$ factor of the duality group is identified with S-duality, one possibility is the presence of $\,\textrm{SL}(2)$-singlet branes (or bound states) in the corresponding ten-dimensional backgrounds. It could be interesting to look at possible uplifts incorporating such branes in a smeared limit. For example, try to add smeared D3-branes to the S-fold setups previously investigated in the literature.  This could shed some light on how to incorporate sources in S-fold backgrounds. Another possibility is that the breaking of supersymmetries in the four-dimensional supergravity Lagrangian (from maximally to half-maximally supersymmetric) is not related to the inclusion of sources but, instead, it stems from geometry. This would be more in the spirit of the half-maximal consistent truncations of \cite{Malek:2017njj} and, perhaps, some generalised frame could be constructed for these half-maximal supergravities along the lines of \cite{Ciceri:2016hup} in order to systematically uplift any four-dimensional solution.

Another interesting line to explore is the possible relation between the AdS$_{4}$ solutions of the $\,\textrm{ISO}(3)_{1} \times \textrm{ISO}(3)_{2}\,$ gaugings of half-maximal supergravity and various classes of type IIB backgrounds of the form $\,\textrm{AdS}_{4} \times \textrm{M}_{6}\,$ that have been constructed directly in ten dimensions using different techniques: pure spinor formalism, G-structures, non-abelian T-duality, ... (see \cite{Lozano:2016wrs,Passias:2017yke,Akhond:2021ffz,Merrikin:2021smb} for an incomplete list). Identifying the field theory duals of these ten-dimensional solutions is a laborious and generically non-systematic task: one first makes an educated guess for the field theory duals and then runs as many holographic tests as possible. Thinking along these lines, it would be very interesting to establish whether or not the general eight-parameter family of $\,\textrm{ISO}(3)_{1} \times \textrm{ISO}(3)_{2}\,$ gaugings of half-maximal supergravity we have presented in Appendix~\ref{app:ISO(3)xISO(3)_general_gauging} describes classes of consistent truncations of type IIB supergravity on $\,\textrm{M}_{6}=\textrm{S}^2 \times \textrm{S}^2 \times \Sigma\,$ with $\,\Sigma\,$ being a Riemann surface. If such a connection exists and is well established via generalised geometry or extended field theory techniques, then exploiting the four-dimensional effective description would provide a way to characterise the CFT$_{3}$'s dual to such type IIB solutions (presumably related to IR fixed points of Gaiotto--Hanany--Witten-like brane constructions) without having to work out their ten-dimensional uplift explicitly. For example, as we have done in this work, the conformal dimensions of the low lying operators in the dual CFT$_{3}$'s could be extracted directly using four-dimensional data, namely, from the mass spectrum of the supergravity fields in the half-maximal $\,\textrm{ISO}(3)_{1} \times \textrm{ISO}(3)_{2}\,$ gauged supergravity. Also the new techniques for Kaluza-Klein (KK) spectrometry put forward in \cite{Malek:2019eaz} could be applied to the type IIB backgrounds of the form $\,\textrm{AdS}_{4} \times \textrm{S}^2 \times \textrm{S}^2 \times \Sigma \,$ (see \cite{Giambrone:2021zvp,Cesaro:2021tna,Giambrone:2021wsm,Cesaro:2022mbu} for a study of the spectrum of KK modes around the type IIB S-folds at $\,\tilde{\varphi}= \pm 1\,$) upon suitable adjustment of the techniques to the context of half-maximal supergravity.

Finally, the analysis performed in Appendix~\ref{app:ISO(3)xISO(3)_general_gauging} shows that the results in the main text can be straightforwardly generalised to include different embedding parameters $\,\tilde{\varphi}_{1,2}\,$ and $\,c_{1,2}$, as well as independent gauge couplings $\,g_{1,2}$, for each of the $\,\textrm{ISO}(3)_{1,2}\,$ factors in the gauge group. In particular, having two independent gauge couplings $\,g_{1,2}\,$ permits to collapse or flatten-out one $\,\textrm{S}^{2}\,$ while keeping the other $\,\textrm{S}^{2}\,$ at finite size. This suggests an a priori much larger structure of supersymmetric AdS$_{4}$ solutions with new potentially interesting CFT$_{3}$ duals. Also going beyond the $\,\mathbb{Z}_{2}^2\,$ and $\,\textrm{U}(1)_{\textrm{R}}\,$ invariant sectors investigated in this work could accommodate new families of AdS$_{4}$ solutions with additional flat directions dual to new marginal deformations in the dual CFT$_{3}$'s. These are all open questions and aspects we plan to continue exploring in the future.

\section*{Acknowledgements}

We are grateful to Ant\'on Faedo, Emanuel Malek, Carlos N\'u\~nez and Daniel Waldram for interesting conversations. The work of AG and MCh-B is partially supported by the Ministry of Science and Universities through the Spanish grant MCIU-22-PID2021-123021NB-I00 and by FICYT through the Asturian grant SV-PA-21-AYUD/2021/52177. CS is supported by IISN-Belgium (convention 4.4503.15) and is a Research Fellow of the F.R.S.-FNRS (Belgium).

\appendix

\section{$\textrm{ISO}(3) \times \textrm{ISO}(3)$ gaugings of half-maximal supergravity}
\label{app:ISO(3)xISO(3)_general_gauging}

In this appendix we analyse the set of possible embeddings of an $\,\textrm{ISO}(3)_{1} \times \textrm{ISO}(3)_{2}\,$ gauging of half-maximal supergravity of the form
\begin{equation}
\label{embedding_chain_appendix}
\textrm{ISO}(3)_{1} \times \textrm{ISO}(3)_{2} \,\,\subset\,\, \textrm{SL}(2) \times \textrm{SO}(3,3)_{1} \times \textrm{SO}(3,3)_{2} \,\,\subset\,\, \textrm{SL}(2) \times \textrm{SO}(6,6) \ .
\end{equation}
We have attached labels $\,_1\,$ and $\,_2\,$ to keep track of each repeated factor. Following the notation of \cite{Dibitetto:2011gm}, and building upon the results of \cite{Roest:2009tt}, a gauging of this type is totally encoded in a set of eight embedding matrices \cite{unpublished}. Since $\,\textrm{SO}(3,3)_{1} \sim \textrm{SL}(4)_{1}\,$, the most general embedding\footnote{This is so up to equivalent solutions of the quadratic constraints in (\ref{QC_N=4}).} of the first $\,\textrm{ISO}(3)_{1}\,$ factor is encoded in four $\,4 \times 4\,$ matrices given by
\begin{equation}
Q_{+}^{(1)} = \left(
\begin{array}{cc}
- a_{0}'  & 0  \\
    0    &  0 \times \mathbb{I}_{3}
\end{array}
\right)
\hspace{8mm} , \hspace{8mm}
\bar{Q}^{(1)}_{+} = \left(
\begin{array}{cc}
0 & 0  \\
    0    &  \tilde{c}_{1}'  \times \mathbb{I}_{3}
\end{array}
\right) \ ,
\end{equation}
\begin{equation}
Q_{-}^{(1)} = \left(
\begin{array}{cc}
- b_{0}'  & 0  \\
    0    &  0 \times \mathbb{I}_{3}
\end{array}
\right)
\hspace{8mm} \, , \hspace{8mm}
\bar{Q}^{(1)}_{-} = \left(
\begin{array}{cc}
0 & 0  \\
    0    &  \tilde{d}_{1}'   \times \mathbb{I}_{3}
\end{array}
\right) \ ,
\end{equation}
with $\,\tilde{c}_{1}' \neq 0\,$. Equivalently, the first $\,\textrm{ISO}(3)_{1}\,$ factor is specified by four embedding tensor components of the form
\begin{equation}
\label{ET_to_fluxes_dictionary_1}
f_{+\bar{a}bc} =   \tilde{c}_{1}' \, \epsilon_{\bar{a}bc}
\hspace{5mm} , \hspace{5mm}
f_{-abc} = - b_{0}' \, \epsilon_{abc}
\hspace{5mm} , \hspace{5mm}
f_{+abc} = - a_{0}' \, \epsilon_{abc}
\hspace{5mm} , \hspace{5mm}
f_{-\bar{a}bc} = \tilde{d}_{1}'  \, \epsilon_{\bar{a}bc} \ .
\end{equation}
Analogously, the second $\,\textrm{ISO}(3)_{2}\,$ factor is encoded in another set of four $\,4 \times 4\,$ matrices of the form
\begin{equation}
Q_{+}^{(2)} = \left(
\begin{array}{cc}
0 & 0  \\
    0    &  \tilde{c}_{2} \times \mathbb{I}_{3}
\end{array}
\right)
\hspace{8mm} , \hspace{8mm}
\bar{Q}^{(2)}_{+} = \left(
\begin{array}{cc}
a_{3}  & 0  \\
    0    &  0 \times \mathbb{I}_{3}
\end{array}
\right) \ ,
\end{equation}
\begin{equation}
Q_{-}^{(2)} = \left(
\begin{array}{cc}
0 & 0  \\
    0    &  \tilde{d}_{2}  \times \mathbb{I}_{3}
\end{array}
\right)
\hspace{8mm} , \hspace{8mm}
\bar{Q}^{(2)}_{-} = \left(
\begin{array}{cc}
b_{3}  & 0  \\
    0    &  0  \times \mathbb{I}_{3}
\end{array}
\right) \ ,
\end{equation}
with $\,\tilde{d}_{2} \neq 0\,$. The components of the embedding tensor for the second $\,\textrm{ISO}(3)_{2}\,$ factor are then given by
\begin{equation}
\label{ET_to_fluxes_dictionary_2}
f_{-i\bar{j} \bar{k}} = \tilde{d}_{2} \, \epsilon_{i\bar{j} \bar{k}}
\hspace{5mm} , \hspace{5mm}
f_{+\bar{i}\bar{j}\bar{k}} = a_{3} \, \epsilon_{\bar{i}\bar{j}\bar{k}}
\hspace{5mm} , \hspace{5mm}
f_{-\bar{i}\bar{j} \bar{k}} = b_{3} \, \epsilon_{\bar{i}\bar{j} \bar{k}}
\hspace{5mm} , \hspace{5mm}
f_{+i\bar{j}\bar{k}} =\tilde{c}_{2} \, \epsilon_{i\bar{j}\bar{k}} \ .
\end{equation}
Together, (\ref{ET_to_fluxes_dictionary_1}) and (\ref{ET_to_fluxes_dictionary_2}) account for all the components of the embedding tensor $\,f_{\alpha MNP}\,$ that are activated in the class (\ref{embedding_chain_appendix}) of $\,\textrm{ISO}(3)_{1} \times \textrm{ISO}(3)_{2}\,$ gaugings of half-maximal supergravity we investigate in this work.

\subsection{Quadratic constraints and algebra structure}

The embedding tensor components in (\ref{ET_to_fluxes_dictionary_1}) and (\ref{ET_to_fluxes_dictionary_2}) automatically satisfy the quadratic constraints of half-maximal supergravity. However, the computation of the additional constraints in (\ref{QC_Extra}) for this multi-parameteric family of $\,\textrm{ISO}(3)_{1} \times \textrm{ISO}(3)_{2}\,$ gaugings yields
\begin{equation}
\label{QC_Extra_general}
f_{\alpha MNP} \, f_{\beta}{}^{MNP}= 0
\hspace{6mm} \textrm{ and } \hspace{6mm}
\left. \epsilon^{\alpha \beta} \, f_{\alpha [MNP} \, f_{\beta QRS]} \right|_{\textrm{SD}}  = 0
\,\, \Leftrightarrow \,\, 
\left\lbrace 
\begin{array}{c}
b_{0}' \, \tilde{c}_{2} - a_{0}' \, \tilde{d}_{2} = 0 \\[2mm]
b_{3} \, \tilde{c}_{1}' - a_{3} \, \tilde{d}_{1}' = 0
\end{array}
 \right. \ .
\end{equation}

The antisymmetry of the commutators $\,[\, T_{\alpha M}\,,\,T_{\beta N}\,] = f_{\alpha MN}{}^{P} \, T_{\beta P}\,$ for this general class of embeddings imposes a set of linear relations between the generators of the form
\begin{equation}
\label{lin_comb_1}
(\tilde{c}_{1}')^{2} \, T_{- a} = \tilde{d}_{1}' \, \tilde{c}_{1}' \,  T_{+a}  + (a_{0}' \, \tilde{d}_{1}' - b_{0}' \, \tilde{c}_{1}') \, T_{+ \bar{a}}
\hspace{8mm} , \hspace{8mm}
\tilde{c}_{1}' \, T_{- \bar{a} }  = \tilde{d}_{1}' \, T_{+ \bar{a}} \ ,
\end{equation}
and
\begin{equation}
\label{lin_comb_2}
(\tilde{d}_{2})^2\, T_{+ \bar{i}} = \tilde{d}_{2} \, \tilde{c}_{2} \,  T_{-\bar{i}} + ( a_{3} \, \tilde{d}_{2} - b_{3} \, \tilde{c}_{2}) \, T_{- i}
\hspace{8mm} , \hspace{8mm}
\tilde{d}_{2} \, T_{+ i }  = \tilde{c}_{2} \, T_{- i} \ .
\end{equation}
Choosing the independent generators to be $\,(T_{+a},T_{+\bar{a}})\,$ and $\,(T_{-i},T_{-\bar{i}})\,$, one finds a set of non-trivial commutation relations of the form
\begin{equation}
\label{algebra_ISO3_1_general}
\begin{array}{llll}
\left[ \,  T_{+ a} , T_{+ b} \, \right] &=&    \tilde{c}_{1}' \, \epsilon_{ab}{}^{c} \,  T_{+ c}  - a_{0}'  \,   \epsilon_{ab}{}^{\bar{c}}  \, T_{+ \bar{c}} \ , \\[4mm]
\left[ \, T_{+ a} , T_{+ \bar{b}} \, \right] &=& \tilde{c}_{1}' \, \epsilon_{a\bar{b}}{}^{\bar{c}} \,  T_{+ \bar{c}} \ , \\[4mm]
\left[ \, T_{+ \bar{a}} , T_{+ \bar{b}} \, \right] &=& 0 \ ,
\end{array}
\end{equation}
for the first ISO(3)$_{1}$ factor in the gauge group and, similarly,
\begin{equation}
\label{algebra_ISO3_2_general}
\begin{array}{llll}
\left[ \,  T_{- \bar{i}} , T_{- \bar{j}} \, \right] &=&     \, \tilde{d}_{2} \,  \epsilon_{\bar{i}\bar{j}}{}^{\bar{k}} \, T_{- \bar{k}}   + b_{3}  \, \epsilon_{\bar{i}\bar{j}}{}^{ k} \, T_{- k}  \ , \\[4mm]
\left[ \, T_{- \bar{i}} , T_{- j} \, \right] &=& \tilde{d}_{2} \, \epsilon_{\bar{i} j}{}^{ k}  \,  T_{- k} \ , \\[4mm]
\left[ \, T_{- i}  ,  T_{- j} \, \right] &=& 0 \ ,
\end{array}
\end{equation}
for the second ISO(3)$_{2}$ factor. Note that, for each of the $\,\textrm{ISO}(3)_{1,2}\,$ factors, there are two parameters entering the commutation relations in (\ref{algebra_ISO3_1_general}) and (\ref{algebra_ISO3_2_general}) and two additional parameters specifying the linear combinations of generators in (\ref{lin_comb_1}) and (\ref{lin_comb_2}).

\subsection{$\mathcal{N}=1\,$ superpotentials}

It is also interesting to investigate the dynamics of the seven moduli fields $\,z_{I}\,$, $\,I=1,\ldots,7\,$, in the $\,\mathbb{Z}_{2}^2$-invariant sector of half-maximal supergravity coupled to six vector multiplets. This sector is described by the $\,\mathcal{N}=1\,$ supergravity multiplet coupled to seven chiral superfields with K\"ahler potential 
\begin{equation}
K = - \sum_{I=1}^{7} \log[- i (z_{I}-\bar{z}_{I}) ] \ ,
\end{equation}
and a superpotential given by
\begin{equation}
\label{W_N=4}
\begin{array}{lll}
 W & = & \Big[  - a_{3}  + a_{0}' \, z_{4} z_{5} z_{6} - \tilde{c}_{2} \, (z_{1} z_{4} + z_{2} z_{5} + z_{3} z_{6}) + \tilde{c}_{1}' \,  (z_{1}z_{5}z_{6}+z_{2}z_{4}z_{6}+z_{3}z_{4}z_{5}) \Big] \\[6mm]
   & + & \Big[\phantom{-} b_{3}  - b_{0}' \, z_{4} z_{5} z_{6} + \tilde{d}_{2} \, (z_{1} z_{4} + z_{2} z_{5} + z_{3} z_{6}) - \tilde{d}_{1}' \,  (z_{1}z_{5}z_{6}+z_{2}z_{4}z_{6}+z_{3}z_{4}z_{5})  \Big] \, z_{7} \ .
\end{array}
\end{equation}
Note that the eight gauging parameters in (\ref{ET_to_fluxes_dictionary_1}) and (\ref{ET_to_fluxes_dictionary_2}) enter the superpotential of the model. 

In this work we have made a simple choice of gauging parameters. More concretely, we have chosen the same embedding for the two $\,\textrm{ISO(3)}_{1,2}\,$ factors in the gauging. This choice drastically simplifies the analysis of supersymmetric vacua. These vacua satisfy the set of supersymmetric (or F-flatness) conditions
\begin{equation}
\label{F=0_eqs_appendix}
F_{I} \equiv D_{I} W =\partial_{I} W + (\partial_{I}K) W = 0 \ ,   
\end{equation}
with $\,I=1, \ldots 7\,$.

\subsection{Back to our model}

The specific model discussed in Section~\ref{sec:Deforming_ISO(3)xISO(3)} corresponds to a simple fixing of the gauging parameters in (\ref{ET_to_fluxes_dictionary_1}) of the form $\, \tilde{d}_{1}' = \tilde{c}_{2} = 0 \,$ and
\begin{equation}
\label{parameters_main_text}
\begin{array}{rcrcrc}
\tilde{c}_{1}' = \frac{2 \, \sqrt{2} \, g }{\sqrt{1 + \tilde{\varphi}^2 }}  &   ,   &   - b_{0}' =  \pm 2 \sqrt{2} \, g \, c \, \frac{\tilde{\varphi}}{\sqrt{1 + \tilde{\varphi}^2 }}     &   ,   &   -a_{0}'  = - 2 \sqrt{2}\, g \, c \,  \frac{\tilde{\varphi}^2-1}{\tilde{\varphi}^2+1} & , \\[6mm]
\tilde{d}_{2}  = \frac{2 \, \sqrt{2} \, g }{\sqrt{1 + \tilde{\varphi}^2 }}   &  ,   & a_{3} =\pm  2 \sqrt{2} \, g \, c \, \frac{\tilde{\varphi}}{\sqrt{1 + \tilde{\varphi}^2 }}   & , & b_{3}  = - 2 \sqrt{2}\, g \, c \,  \frac{\tilde{\varphi}^2-1}{\tilde{\varphi}^2+1}  & .
\end{array}
\end{equation}
Plugging (\ref{parameters_main_text}) into (\ref{W_N=4}) yields a superpotential
\begin{equation}
\label{W_N=4_varphitilde_1}
\begin{array}{lll}
W &=& \dfrac{2 \, \sqrt{2} \, g}{\sqrt{1+\tilde{\varphi}^2}} \, \Big[z_{1}z_{5}z_{6}+z_{2}z_{4}z_{6}+z_{3}z_{4}z_{5}  + (z_{1} z_{4} + z_{2} z_{5} + z_{3} z_{6}) \, z_{7}\Big] 
\\[4mm]
& - & \dfrac{2 \, \sqrt{2} \, g}{\sqrt{1+\tilde{\varphi}^2}} \, c \, \Big[ \pm \tilde{ \varphi} \, (1- z_{4}z_{5}z_{6}z_{7})  + \dfrac{1 - \tilde{\varphi}^2}{\sqrt{1+\tilde{\varphi}^2}} (z_{4}z_{5}z_{6} - z_{7})\Big] \ .
\end{array}
\end{equation}
We have verified that the AdS$_{4}$ solutions in (\ref{vacuum_N=4_N=2}) solve the F-flatness equations in (\ref{F=0_eqs_appendix}) constructed from (\ref{W_N=4_varphitilde_1}).

Lastly, as a further check of consistency, setting $\,\tilde{\varphi}^2 = 1\,$ recovers the maximal theory. Namely, the superpotential in (\ref{W_N=4_varphitilde_1}) reduces to
\begin{equation}
\label{W_N=8}
W = 2 \, g \, \Big[z_{1}z_{5}z_{6}+z_{2}z_{4}z_{6}+z_{3}z_{4}z_{5}  + (z_{1} z_{4} + z_{2} z_{5} + z_{3} z_{6}) \, z_{7}\Big] \pm 2 \, g \, c \, \left( 1 - z_{4} z_{5} z_{6} z_{7}  \right) \ ,
\end{equation}
in agreement with the result of \cite{Guarino:2020gfe}.

\bibliography{references}
\end{document}